# Approximability and proof complexity

Ryan O'Donnell[*] Yuan Zhou[†]

November 9, 2012

**Abstract**

This work is concerned with the proof-complexity of certifying that optimization problems do *not* have good solutions. Specifically we consider bounded-degree "Sum of Squares" (SOS) proofs, a powerful algebraic proof system introduced in 1999 by Grigoriev and Vorobjov. Work of Shor, Lasserre, and Parrilo shows that this proof system is automatizable using semidefinite programming (SDP), meaning that any $n$-variable degree-$d$ proof can be found in time $n^{O(d)}$. Furthermore, the SDP is dual to the well-known Lasserre SDP hierarchy, meaning that the "$d/2$-round Lasserre value" of an optimization problem is equal to the best bound provable using a degree-$d$ SOS proof. These ideas were exploited in a recent paper by Barak et al. (STOC 2012) which shows that the known "hard instances" for the Unique-Games problem are in fact solved close to optimally by a constant level of the Lasserre SDP hierarchy.

We continue the study of the power of SOS proofs in the context of difficult optimization problems. In particular, we show that the Balanced-Separator integrality gap instances proposed by Devanur et al. can have their optimal value certified by a degree-$4$ SOS proof. The key ingredient is an SOS proof of the KKL Theorem. We also investigate the extent to which the Khot–Vishnoi Max-Cut integrality gap instances can have their optimum value certified by an SOS proof. We show they can be certified to within a factor $.952$ ($> .878$) using a constant-degree proof. These investigations also raise an interesting mathematical question: is there a constant-degree SOS proof of the Central Limit Theorem?

[*]Department of Computer Science, Carnegie Mellon University. Supported by NSF grants CCF-0747250 and CCF-1116594, a Sloan fellowship, and a grant from the MSR–CMU Center for Computational Thinking.

[†]Department of Computer Science, Carnegie Mellon University. Partially supported by a grant from the Simons Foundation (Award Number 252545).

## 1 Introduction

In a typical constraint satisfaction problem (CSP) we are given a set of variables $V$ to be assigned values from some finite domain $\Omega$ (often $\{0,1\}$); we are also given a set of local constraints specifying how various small groups of variables should be assigned. The task is to find an assignment to the variables which minimizes the number of unsatisfied constraints. Sometimes there may also be inviolable global constraints; for example, that no domain element is assigned to too many variables. A canonical example is the Balanced-Separator problem: given is a graph $(V, E)$ with $n$ vertices which must be partitioned into two "balanced" parts, each of cardinality at least $n/3$; the goal is to minimize the number of edges crossing the cut.

For such problems, certifying that there is a good solution is in NP; for example, given a graph we can efficiently prove that it has a balanced cut of size at most $\alpha$ simply by exhibiting the cut. But what about the opposite problem, certifying that every balanced cut has size at least $\beta$? Since this problem is coNP-complete it is unlikely that there are efficient certifications for every instance; however there may be efficient certifications for specific instances or classes of instances. For example, if we consider a linear programming relaxation of a given Balanced-Separator instance and then exhibit a dual solution of value $\beta$, this constitutes a proof that every balanced cut in the instance has size at least $\beta$.

The question is also interesting for problems in P, especially when the complexity of the proof system is taken into account. For example, given an unsatisfiable instance $Ax = b$ of the 3Lin2 CSP (meaning the equations are over $\mathbb{F}_2$ and each involves at most 3 variables), there is always an easy-to-verify proof of unsatisfiability: a vector $y$ such that $y^\top A = 0$ but $y^\top b \neq 0$. However finding such a proof requires a rather specialized algorithm, Gaussian Elimination. By contrast, unsatisfiable instances of the 2Lin2 CSP have simple proofs of unsatisfiability (an unsatisfiable "cycle" of variables) which can be found by a very generic "local consistency" algorithm. Indeed, one can view this algorithm as searching for all constant-width Resolution proofs of unsatisfiability; the same algorithm works for any "bounded-width CSP" [2].

**Positivstellensatz proofs.** In this work we consider a certain strong proof system for CSPs. It belongs to the well-studied class of *algebraic proof systems*, in which local constraints are represented by polynomial equations. To handle global constraints we also allow for polynomial *inequalities*; this is also natural in the context of the linear programs and semidefinite programs used by optimization algorithms. To give an example, suppose we have a Balanced-Separator instance $(V, E)$ with $V = [n]$. We introduce a real variable $X_i$ for each $i \in V$. Now to say that the optimum value of the instance is larger than $\beta$ is precisely equivalent to saying the following system of polynomial equations and inequalities (each of degree at most 2) is *infeasible*:

$$A = \Big\{X_i^2 = X_i \ \forall i \in [n]\Big\} \cup \Big\{\sum_{i=1}^n X_i \geq n/3,\ \sum_{i=1}^n X_i \leq 2n/3\Big\} \cup \Big\{\sum_{(i,j)\in E} (X_i - X_j)^2 \leq \beta\Big\}.$$

Here the first set of equations enforces $X_i \in \{0,1\}$, encoding a cut. The second set of inequalities enforces that the cut is balanced. The final inequality states that at most $\beta$ edges cross the cut. Now what would constitute a proof that $A$ has no real solutions; i.e., that the Balanced-Separator value exceeds $\beta$? One certificate would be a formal identity in the polynomial ring $\mathbb{R}[X_1, \dots, X_n]$ of the following form:

$$-1 = \sum_{i=1}^n P_i \cdot (X_i^2 - X_i) + U \cdot (\sum_{i=1}^n X_i - n/3) + U' \cdot (2n/3 - \sum_{i=1}^n X_i) + V \cdot (\beta - \sum_{(i,j)\in E} (X_i - X_j)^2) + W, \quad (1)$$

where $P_1, \dots, P_n \in \mathbb{R}[X_1, \dots, X_n]$ and where $U, U', V, W \in \mathbb{R}[X_1, \dots, X_n]$ are each *sums of squares (SOS)*, meaning of the form $Q_1^2 + Q_2^2 + \cdots + Q_m^2$ for some $Q_1, \dots, Q_m \in \mathbb{R}[X_1, \dots, X_n]$. Such an identity would indeed imply that $A$ is infeasible, since substituting any solution of $A$ into (1) would give a nonnegative right-hand side.

In fact, a certain refinement [49] of the *Positivstellensatz* of Krivine [32] and Stengle [55] guarantees that if $A$ is infeasible then there is *always* a proof of the form (1). A generic "SOS proof system" based on the Positivstellensatz was introduced around 1999 by Grigoriev and Vorobjov [19]. As with most algebraic proof systems it can be difficult to place an a priori upper bound on the *degree* of the polynomials needed for a proof; if we insist on a fixed degree bound $d$ then the proof system becomes incomplete. On the other hand this incomplete system has the advantage of being efficiently *automatizable*, meaning that if a proof exists it can be found in time $\mathrm{poly}(n^d)$. The algorithm uses semidefinite programming and follows from the work of Shor [54], Nesterov [44], Lasserre [33, 34] and Parrilo [47]. See Section 1.1 for more details.

**The power of SOS.** Most of the previous relevant work focused on showing SOS-degree (equivalently, Lasserre-round) lower bounds. However a recent paper by Barak, Brandão, Harrow, Kelner, Steurer, and Zhou [3] brought to light the importance of SOS degree *upper* bounds for the study CSP approximability. That paper considered the strong integrality gap instances known for the notorious Unique-Games CSP [50, 27, 4] and (essentially) showed that degree-8 SOS proofs can certify that the instances have value close to $0$. Thus the generic $\mathrm{poly}(n)$-time "level-4 Lasserre SDP" algorithm refutes their having large optimal value. This is despite the fact that the instances still have value near $1$ after $\Theta(\log\log n)^{1/4}$ rounds of the rather powerful Sherali–Adams SDP hierarchy [50].

The purpose of the present paper is to further explore the relevance of SOS proof complexity to the algorithmic theory of CSP approximation. Specifically, we show that the Devanur–Khot-Saket–Vishnoi [12] instances of Balanced-Separator can have their optimal value well-certified by a degree-$4$ SOS proof. We also investigate the problem of SOS proofs for the Khot–Vishnoi (KV) [29] instances of Max-Cut and raise an intriguing mathematical question: is there an SOS proof of the Central Limit Theorem?

## 1.1 History

We review here some of what is known about SOS proofs and SDP hierarchies; for a much more thorough discussion we recommend the monograph by Laurent [38].

Throughout this work we write $X = (X_1, \dots, X_n)$ for a sequence of indeterminates, with the number $n$ being clear from context. We say that the real multivariate polynomial $u \in \mathbb{R}[X]$ is *sum of squares (SOS)* if $u = s_1^2 + \cdots + s_m^2$ for some $s_1, \dots, s_m \in \mathbb{R}[X]$. Any SOS polynomial is nonnegative on all of $\mathbb{R}^n$; however, as Hilbert [21] showed in 1888 there exist nonnegative polynomials which are not SOS. The first explicit example, $X_1^2X_2^2(X_1^2+X_2^2-3)+1$, was given by Motzkin in the mid-'60s. Hilbert's 17th Problem [22] asks whether every nonnegative polynomial $q$ is the quotient of SOS polynomials; this was solved affirmatively by Artin [1].

Artin's result also follows from the Positivstellensatz, first proved (essentially) by Krivine [32] and then later independently by Stengle [55]. Interestingly, Stengle's motivation was the duality theory of linear programming. We state a special case appearing in [6]:

**Positivstellensatz.** *Let $A$ be a finite set of real multivariate polynomial equations and inequalities,*

$$A = \{p_1 = 0, p_2 = 0, \dots, p_m = 0\} \cup \{q_1 \geq 0, q_2 \geq 0, \dots, q_{m'} \geq 0\},$$

*with each $p_i, q_j \in \mathbb{R}[X]$. Then $A$ is infeasible if and only if there exist polynomials $r_1, \dots, r_m$ and SOS polynomials $(u_J)_{J\subseteq[m']}$ in $\mathbb{R}[X]$ such that*

$$-1 = \sum_{i=1}^{m} r_i p_i + \sum_{J\subseteq[m']} u_J \prod_{j\in J} q_j. \tag{2}$$

One interesting further special case occurs when $A$ contains only equations, not inequalities. In this case the Positivstellensatz says that $p_1, \dots, p_m$ have no common real roots if and only if the ideal they generate contains $1 + u$ for some SOS $u$. This special case arises whenever one wants to show that a CSP (with no global constraints) is not perfectly satisfiable. (As noted by Shor [53], one can actually reduce to this case in general by replacing $q \geq 0$ with $q - Y^2 = 0$, where $Y$ is a new indeterminate; indeed, by further substitutions of new indeterminates one can reduce to the case where all equations are quadratic.)

**Proof complexity.** Extending the Nullstellensatz proof system of Beame, Impagliazzo, Krajíček, Pitassi, and Pudlák [5], Grigoriev and Vorobjov [19] proposed in 1999 the natural propositional proof system based on the Positivstellensatz. The complexity measure is degree: i.e., $\max_{i,J}\{r_i p_i, u_J \prod_{j \in J} q_j\}$ in (2). This is a *static* proof system, meaning that one simply exhibits the refutation (2).[1] Grigoriev and Vorobjov showed that refuting the single equation

$$(1 - X_0 X_1)^2 + (X_1^2 - X_2)^2 + (X_2^2 - X_3)^2 + \cdots + (X_{n-1}^2 - X_n)^2 + X_n^2 = 0$$

requires a proof of degree at least $2^{n-1}$. Relying on some ideas from the work of Buss, Grigoriev, Impagliazzo, and Pitassi [9], Grigoriev showed in 1999 [15, 17] that refuting any unsatisfiable system of $\mathbb{F}_2$-linear equations requires degree at least $D/2$, where $D$ is the least width needed to give a Resolution refutation. As a consequence he showed that degree $\Omega(n)$ is necessary to prove Tseitin tautologies on $n$-vertex regular expander graphs and to prove that the graph $K_n$ has no perfect matching when $n$ is odd. Grigoriev also subsequently [16] showed that the "$r$-Knapsack tautology" requires a proof of degree $n + 1$ for any real $r \in (\frac{n}{2} - \frac{1}{2}, \frac{n}{2} + \frac{1}{2})$; this is the infeasibility of the system

$$\{X_1^2 = X_1, \dots, X_n^2 = X_n, X_1 + \cdots + X_n = r\},$$

for $r$ a non-integer. For more on algebraic proof complexity with inequalities, see e.g. [48, 18].

**Optimization.** We now discuss algorithmic issues. Let $u \in \mathbb{R}[x]$ be a real $n$-variate polynomial of degree $d$. A most basic optimization problem is to determine $\inf_{x \in \mathbb{R}^n} u(x)$. Roughly speaking, this is equivalent (by binary search) to the problem of deciding whether $u(x) \geq \alpha$; further, there is no loss of generality in assuming $\alpha = 0$. Unfortunately, the problem of deciding whether $u \geq 0$ is $\mathsf{NP}$-hard as soon as $d \geq 4$. In 1987, Shor [54] pioneered the idea of replacing the condition $u \geq 0$ with the stronger condition that $u$ is SOS, and noted that this can be tested in $\mathrm{poly}(n^d)$ by solving an SDP feasibility problem. (Here we ignore the issue of precision in solving SDPs; see Section 2 for more details.) Shor made the connection to Hilbert's 17th Problem but not to Positivstellensatz.

Beginning in 2000, Parrilo [47] and Lasserre [33, 34] independently published several works taking the idea further. Parrilo emphasized the viewpoint of Positivstellensatz as a refutation system for polynomial inequalities, while Lasserre focused significant attention on the dual SDP "problem of moments". Both proposed using $\mathrm{poly}(n^d)$-time SDPs to search for degree-$d$ Positivstellensatz refutations, for larger and larger $d$.

Lasserre also proposed using certain variant forms of Positivstellensatz. For example, if one is optimizing a polynomial on a *compact* semialgebraic set $K$ then one can use SDP optimization directly (as opposed to using binary search and feasibility testing), thanks to a version of the Positivstellensatz due to Schmüdgen [51]. Furthermore, Putinar [49] showed that if $K$ is explicitly compact ("Archimedean") — say, one of its defining inequalities is $\sum_{i=1}^n X_i^2 \leq B$ — then the Positivstellensatz certificates (2) only require $u_J$'s with $|J| \leq 1$. (Both [51, 49] contained a bug, fixed in [57].) On one hand, in practice there is rarely any harm in adding an inequality $\sum_{i=1}^n X_i^2 \leq B$ with large $B$; on the other hand, eliminating the $u_J$'s with $|J| > 1$ may cause the refutation degree to increase. In any case, Lasserre focused on the polynomial optimization problem

$$\inf\{p(x) \mid x \in K\}, \quad K = \{x \in \mathbb{R}^n \mid q_1(x) \geq 0, \dots, q_m(x) \geq 0\}, \tag{3}$$

and proposed a hierarchy of SDP relaxations for increasing $d$,

$$\inf\{L(p) \mid L : \mathbb{R}[X]_d \to \mathbb{R} \text{ is a linear map}, L(1) = 1, \text{ and } L(u), L(uq_i) \geq 0 \text{ for all SOS } u\}, \tag{4}$$

where $\mathbb{R}[X]_d$ denotes the ring $\mathbb{R}[X]$ restricted to polynomials of degree at most $d$. This is a relaxation because one can take $L$ to be the evaluation map $p \mapsto p(x^*)$ for any optimal solution $x^*$. (In [3], $L(p)$ is written as $\widetilde{\mathbf{E}}[p]$ and termed the "pseudo-expectation" of $p$.) We refer to (4) as the *degree-$d$ Lasserre moment SDP*; when $d$ is even it is also known as the *level-$d/2$* (or sometimes $d/2-1$) *Lasserre hierarchy SDP*. The semidefinite dual of (4) is

$$\sup\{\beta \mid p - \beta = u_0 + u_1q_1 + \cdots + u_mq_m \text{ for some SOS } u_0, \dots, u_m \text{ with } \deg(u_0), \deg(u_iq_i) \leq d\}, \tag{5}$$

which we refer to as the *degree-$d$ Lasserre SOS SDP*. (One can also allow for polynomial equalities in the description of $K$, either by replacing them with pairs of inequalities, extending the SDP formulations as in (2), or by factoring out by the ideal they generate [37].) Assuming $K$ is explicitly compact, Lasserre [34] showed that the SOS SDP's value tends to the optimal value as the degree increases. If furthermore $K$ has a nonempty interior then there is no duality gap between (4) and (5). Generally $K$ has *empty* interior for discrete optimization problems (e.g., if it includes the constraints $X_i^2 = X_i$); however, the duality gap issue is algorithmically irrelevant since the Ellipsoid Algorithm can't distinguish an empty interior from a small interior anyway. This issue is discussed briefly in Section 2.

[1] Grigoriev and Vorobjov also proposed a certain dynamic version of the proof system, analogous to Polynomial Calculus [11]. Indeed, [39] had earlier proposed a dynamic proof system based on Positivstellensatz. We do not discuss dynamic proof systems further in this paper.

**Prior optimization results.** We conclude by mentioning some known positive and negative results for the Lasserre moment SDP relaxation. Around 2001, Laurent [36] considered the Lasserre hierarchy for Max-Cut with negative edge weights allowed (i.e., the 2Lin2 CSP). She showed that degree-2 Lasserre optimally solves all instances whose underlying graph is a tree, and conversely that there are non-tree instances which degree-2 Lasserre does not solve optimally. She similarly characterized the underlying graphs which degree-4 Lasserre solves optimally: the $K_5$-minor-free graphs. Around 2002, Laurent [35] showed that when $n$ is odd, the degree-$(n-1)$ moment SDP relaxation for the Max-Cut problem on $K_n$ still has value $\frac{n^2}{4}$ (whereas the optimum value is $\frac{n^2-1}{4}$); i.e., the $\lceil\frac{n+1}{2}\rceil^{\text{th}}$ level of the Lasserre hierarchy is required to obtain the optimal solution. Around 2005, Cheung [10] considered the Knapsack problem and showed that in the optimization problem

$$\inf\{X_1 + \cdots + X_n \mid X_i^2 = X_i \; \forall i, \; X_1 + \cdots + X_n \geq r\},$$

if $r = r(n) \in (0, 1)$ is sufficiently small then the Lasserre moment SDP does not find the optimal solution (namely, 1) until the degree is "maximal", namely $2n + 2$. In 2008, Schoenbeck essentially rediscovered Grigoriev's result on $\mathbb{F}_2$-linear equations from the moment side, showing that there are $n$-variable 3Lin2 instance of value $\frac{1}{2}+o_n(1)$ for which the degree-$\Omega(n)$ Lasserre moment relaxation still has value 1. Building on this work, Tulsiani [56] showed degree-$\Omega(n)$ integrality gap instances matching the known NP-hardness factors for a number of CSPs. Guruswami, Sinop, and Zhou [20] showed a degree-$\Omega(n)$ integrality gap instance for the Balanced-Separator problem with factor $\alpha > 1$, even though this level of NP-hardness is not known. They also showed a degree-$\Omega(n)$ integrality gap instance for the Max-Cut problem with factor $\frac{17}{18}$. Around 2010, Karlin, Mathieu, and Nguyen [24] showed that the degree-$2t$ Lasserre moment relaxation achieves approximation ratio $1 - \frac{1}{t}$ for the general Knapsack problem.

## 1.2 Our contributions: an outline of this work

Continuing a line of work begun in [3], we investigate whether the $O(1)$-degree SOS SDP hierarchy can solve known integrality gap instances of problems that are essentially *harder* than Unique-Games. We focus on two such problems: Balanced-Separator and Max-Cut.

**Balanced-Separator.** Building on work of Khot–Vishnoi [29] and Krauthgamer–Rabani [31], Devanur, Khot, Saket, and Vishnoi (DKSV) [12] gave a family of $n$-vertex Balanced-Separator instances in which the optimal balanced separator cuts an $\Omega(\frac{\log\log n}{\log n})$ fraction of the edges, but for which the SDP with triangle inequalities has value $O(\frac{1}{\log n})$. This is a factor-$\Theta(\log\log n)$ integrality gap. Raghavendra and Steurer [50] show that a factor-$(\log\log n)^{\Omega(1)}$ gap persists for these instances even for $(\log\log n)^{\Omega(1)}$ rounds of the "LH SDP hierarchy". The key to analyzing the optimum value of their instances is the KKL Theorem [23] from analysis of boolean functions. In this work we give a degree-4 SOS proof of the KKL Theorem. In turn, this is used in Section 6 to show the following:

**Theorem.** *The degree-4 SOS relaxation for the DKSV Balanced-Separator instances has value* $\Omega(\frac{\log\log n}{\log n})$.

Thus just the level-2 Lasserre SDP hierarchy (essentially) solves the DSKV Balanced-Separator instances.

**Max-Cut.** Khot and -Vishnoi [29] gave integrality gap instances for the Max-Cut problem, by composing their Unique-Games instances with the Khot–Kindler–Mossel–O'Donnell [26] Max-Cut reduction. When this reduction is executed with parameter $\rho \in (-1, 0)$, one obtains $n$-vertex Max-Cut instances with optimal value at most $(\arccos\rho)/\pi + o_n(1)$, but for which the SDP with triangle inequalities has value $\frac{1}{2} - \frac{1}{2}\rho - o_n(1)$. In particular, for $\rho = \rho_0 \approx -.689$, this is a factor-$.878$ integrality gap (worst possible, by the Goemans–Williamson algorithm [14]). Khot and Saket [28] subsequently showed that this gap persists even for $(\log\log\log n)^{\Omega(1)}$ rounds of the Sherali–Adams SDP hierarchy. The key to analyzing the optimum value of the KV Max-Cut instances is the Majority Is Stablest Theorem from [42]. This theorem is in turn based on an Invariance Principle for nonlinear forms of random variables, together with a Gaussian isoperimetric theorem of Borell [8]. We are able to "SOS-ize" Kindler–O'Donnell's recent new proof of the latter [30] (it essentially only needs the triangle inequality); however we do not know how to prove the former for non-polynomial functionals. Thus we currently do not know how to give an SOS proof of the Majority Is Stablest Theorem.

We turn then to a weaker version of Majority Is Stablest known as the "$\frac{2}{\pi}$ Theorem", proved in [25]. This proof relies on just the Central Limit Theorem (more precisely, the Berry–Esseen Theorem). We *are* able to give an SOS proof of the CLT Theorem, although not with a fixed constant degree bound. Rather, we are able to prove it up to an additive error of $\delta$ using an SOS proof of degree $\widetilde{O}(1/\delta^2)$. Using this, as well as the SOS analysis of the KV Unique-Games instances due to [3], we are able to show the following in Section 8:

**Theorem.** *There exists a universal constant $C \in \mathbb{N}^+$ such that the degree-$C$ SOS relaxation for the KV Max-Cut instances (with parameter $\rho_0 \approx -.689$) is within a factor $.952$ ($> .878$) of the optimum value. For general $\rho$, the relaxation is within a factor of $.931$ of the optimum.*

**A guide to the SOS proofs.** Since even conceptually simple SOS proofs can sometimes look a little complicated, we give here a brief guide to our SOS proofs. Both of our results rely on the hypercontractive inequality for $\{-1, 1\}^n$ due to [7]. Barak et al. [3] already gave a degree-4 SOS proof of one form of this inequality. The only trick is that to evade the use of Cauchy–Schwarz in the standard proofs one needs to move to a "two-function" version of the inequality. We need SOS proofs of a few other forms of the hypercontractive inequality, which we provide in Section 4. Though the notation is heavy, the proofs are essentially straightforward. On the other hand, we remark that we currently do not have an SOS proof of the $2 \to 2k$ version of the inequality with sharp constant for any integer $k > 2$.

In KKL, hypercontractivity is used to prove the "Small-Set Expansion (SSE) in the Noisy Hypercube" theorem. The usual proof of this is very short, but presents a couple of challenges for SOS proofs. One challenge is the use of Hölder's inequality with exponents $4, \frac{4}{3}$. We are able to get around the fractional powers with a couple of tricks, one which is the following: if one needs to SOS-prove, say, $p \leq \sqrt{q}$ for some nonnegative polynomial $q$, instead prove that $p \leq \frac{\epsilon}{2} + \frac{1}{2\epsilon} \cdot q$ for all real $\epsilon > 0$. The other challenge is that the standard proof of the SSE Theorem involves division by a polynomial quantity, something we don't see how to do with SOS proofs. Still, we manage to give a short SOS-proof of a weaker version of the SSE Theorem which is good enough for our purposes. We remark that our proof is somewhat similar to the Barak et al. analysis of the KV Unique-Games instances; however because we work on the SOS side rather than the moment side, we need a few extra tricks. Finally, to obtain the Balanced-Separator result, the last step is to SOS-prove the KKL Theorem. Even the statement of the theorem involves logarithms, which does not look SOS-friendly. We get around this with a variant of the square-root trick just mentioned.

Moving to our proof of the $\frac{2}{\pi}$ Theorem, as stated, we need an SOS-proof of the Central Limit Theorem (with error bounds). Alternately phrased, we need an Invariance Theorem for linear forms of polynomials, specifically with the absolute-value functional. Although this functional is not polynomial, we can replace the required statement with something that is: namely, when $a_1, \dots, a_n$ are indeterminates assumed to satisfy $a_1^2 + \cdots + a_n^2 = 1$, we want to upper-bound

$$\mathop{\mathbf{E}}_{\boldsymbol{x} \sim \{-1,1\}^n}[f(\boldsymbol{x})(a_1\boldsymbol{x}_1 + \cdots + a_n\boldsymbol{x}_n)] \leq \sqrt{\frac{2}{\pi}} + e,$$

where $e$ is an error term involving $\sum_i a_i^4$, which is small when all $a_i$'s are small. Our SOS proof of this is somewhat technically difficult. To proceed, we upper-bound the absolute-value functional to within $\delta$ by a polynomial $Q$ of high degree; using real approximation theory, $\widetilde{O}(1/\delta^2)$ suffices. Then we prove an Invariance Theorem for linear forms with a high-degree functional; this is feasible for linear forms (but not higher-degree ones) because of their subgaussian tails. Unlike in the usual proof of the Berry–Esseen Theorem, we need the hypercontractive inequality for high norms here.

## 2 The SOS proof system and the SDP hierarchy for optimization

In this section we give formal details of the Positivstellensatz proof system of Grigoriev–Vorobjov and the associated hierarchy of SDP algorithms due to Lasserre and Parrilo. For brevity we refer to these as "SOS proofs and hierarchies".

**Definition 2.1.** Let $X = (X_1, \dots, X_n)$ be indeterminates, let $q_1, \dots, q_m, r_1, \dots, r_{m'} \in \mathbb{R}[X]$, and let

$$A = \{q_1 \geq 0, \dots, q_m \geq 0\} \cup \{r_1 = 0, \dots, r_{m'} = 0\}.$$

Given $p \in \mathbb{R}[X]$ we say that $A$ *SOS-proves* $p \geq 0$ *with degree* $k$, written

$$A \quad \vdash_k \quad p \geq 0,$$

whenever

$\exists v_1, \dots, v_{m'}$ and SOS $u_0, u_1, \dots, u_m$ such that

$$p = u_0 + \sum_{i=1}^{m} u_i q_i + \sum_{j=1}^{m'} v_j r_j, \qquad \text{with } \deg(u_0), \deg(u_i q_i), \deg(v_j r_j) \leq k \ \forall i \in [m], j \in [m'].$$

(Recall we say that $w \in \mathbb{R}[X]$ is SOS if $w = s_1^2 + \cdots + s_t^2$ for some $s_i \in \mathbb{R}[X]$.)

We say that $A$ has a *degree-$k$ SOS refutation* if

$$A \quad \vdash_k \quad -1 \geq 0.$$

Finally, when $A = \emptyset$ we will sometimes use the shorthand

$$\vdash_k \quad p \geq 0,$$

which simply means that $p$ is SOS and $\deg(p) \leq k$.

Our notation here is suggestive of a dynamic proof system, and indeed it can be helpful to think of SOS proofs this way. For example, adding deductions is not a problem:

**Fact 2.2.** *If*

$$A \quad \vdash_k \quad p \geq 0, \qquad A' \quad \vdash_{k'} \quad p' \geq 0,$$

*then*

$$A \cup A' \quad \vdash_{\max(k,k')} \quad p + p' \geq 0.$$

However using transitivity or multiplying together two deductions leads to a worse degree bound when applied generically:

**Fact 2.3.** *Suppose that*

$$A \quad \vdash_k \quad q_1' \geq 0, \ \ldots, \ q_\ell' \geq 0$$

*(meaning $A \vdash_k q_i' \geq 0$ for each $i \in [\ell]$). Further suppose that*

$$\{q_1' \geq 0, \ldots, q_\ell' \geq 0\} \quad \vdash_{k'} \quad p \geq 0.$$

*Then*

$$A \quad \vdash_{k+k'} \quad p \geq 0.$$

**Fact 2.4.** *Let $A = \{q_1 \geq 0, \ldots, q_m \geq 0\}$, $A' = \{q_1' \geq 0, \ldots, q_{m'}' \geq 0\}$. If*

$$A \quad \vdash_k \quad p \geq 0, \qquad A' \quad \vdash_{k'} \quad p' \geq 0,$$

*then*

$$A \cup A' \cup (A \cdot A') \quad \vdash_{k+k'} \quad p \cdot p' \geq 0,$$

*where $A \cdot A'$ denotes $\{q_i \cdot q_j' \geq 0 : i \in [m], j \in [m']\}$.*

Notice that in the above fact we had to explicitly include product inequalities into the hypotheses. This is because in general we do not have $\{q \geq 0, q' \geq 0\} \vdash qq' \geq 0$. For example:

**Proposition 2.5.** *In $\mathbb{R}[Y, Z]$, for every $k \in N$,*

$$\{Y \geq 0, Z \geq 0\} \quad \nvdash_k \quad YZ \geq 0.$$

*Indeed, for all real $\beta \geq 0$,*

$$\{Y \geq 0, Z \geq 0\} \quad \nvdash_k \quad YZ \geq -\beta.$$

*Proof.* Suppose to the contrary that

$$YZ + \beta = u_1 + Yu_2 + Zu_3 \tag{6}$$

for some SOS $u_1, u_2, u_3 \in \mathbb{R}[Y, Z]$. We think of the right-hand side of (6) as being in $\mathbb{R}[Z][Y]$. Let $k_j$ be the degree of $Y$ in $u_j$ for $j = 1, 2, 3$; note that $k_1, k_3$ are even and $k_2$ is odd. Suppose first that $\max\{k_1, k_2, k_3\} = k_1$. Then we must in fact have $k_3 = k_1$ in order to cancel the $Y^{k_1}$ term in the RHS of (6). But in fact such a cancelation is impossible because the coefficient on $Y^{k_1}$ in $u_1$ will be an even-degree polynomial in $Z$, but the coefficient on $Y^{k_3}$ in $u_3$ will be an odd-degree polynomial in $Z$. The remaining possibility is that $k_2 > k_1, k_3$. In this case we must have $k_2 = 1$, or else the degree of $Y$ on the RHS of (6) will exceed 1. Thus $u_1, u_2, u_3$ depend only on $Z$; but then (6) forces $u_2 = Z$, contradicting the fact that $u_2$ is SOS. □

For more simple examples of the weakness of SOS proofs, see [41, Chap. 2.7]. Here is another one: we cannot directly prove $Y^4 \geq 1 \Rightarrow Y^2 \geq 1$.

**Proposition 2.6.** *In $\mathbb{R}[Y]$, for every $k \in N$,*

$$Y^4 \geq 1 \quad \not\vdash_k \quad Y^2 \geq 1.$$

*Proof.* Suppose to the contrary that one can write

$$Y^2 - 1 = u + v(Y^4 - 1) \tag{7}$$

with $u, v \in \mathbb{R}[Y]$ being SOS. One cannot have $v = 0$ because $Y^2 - 1$ is not SOS (consider that $0^2 - 1$ is negative). Therefore the highest-degree term in $v$ is of the form $cY^{2j}$ for some real $c > 0$ and some integer $j$. This gives a term $cY^{2j+4}$ on the right-hand side of (7) which must be canceled by $u$. This is impossible if $\deg(u) = 2j + 4$ because the leading coefficient on $u$ will be positive too. So $\deg(u) > 2j + 4$, but then *its* highest-degree term remains uncanceled on the right-hand side of (7). □

On the other hand, one *can* easily SOS-prove $Y^4 \leq 1 \Rightarrow Y^2 \leq 1$; see Fact 3.3. Furthermore, one can $Y^4 \geq 1 \Rightarrow Y^2 \geq 1$ by contradiction:

**Proposition 2.7.** *In $\mathbb{R}[Y]$, for any $\epsilon > 0$ we have*

$$\{Y^4 \geq 1, Y^2 \leq 1 - \epsilon\} \quad \vdash_4 \quad -1 \geq 0.$$

*Proof.* We leave the case of $\epsilon \geq 1$ to the reader. Otherwise, write $c = 1 - \epsilon \in (0, 1)$; then

$$-1 = \tfrac{1}{1-c^2}(c + Y^2)(c - Y^2) + \tfrac{1}{1-c^2}(Y^4 - 1)$$

and both $\frac{1}{1-c^2}(c + Y^2)$ and $\frac{1}{1-c^2}$ are SOS. □

These observations reveal that when fixing the degree of SOS proofs, the SDP simplifications explored by Lasserre (see Section 1.1) can be damaging: it may help to multiply together constraint inequalities, and direct optimization can be worse than binary searching for refutations. Thus we propose that for optimizations problems, one should generically use the SDP hierarchy proposed by Parrilo. I.e., for

$$\inf\{p(x) \mid x \in K\}, \qquad K = \{x \in \mathbb{R}^n \mid q_1(x) \geq 0, \ldots, q_m(x) \geq 0\},$$

one should assume that $K$ is "explicitly compact" (say, contains the inequality $X_1^2 + \cdots + X_n^2 \leq 2^{\mathrm{poly}(n)}$) and then use binary search to (approximately) find the largest $\beta$ for which

$$\{q_{i_1} q_{i_2} \cdots q_{i_t} \geq 0 : \deg(q_{i_1} q_{i_2} \cdot \cdots \cdot q_{i_t}) \leq d\} \cup \{p \leq \beta\} \quad \vdash_d \quad -1 \geq 0. \tag{8}$$

This can be carried out in $\mathrm{poly}(n^d, m)$ time using the Ellipsoid Algorithm.[2]

[2]Determining (8) amounts to checking if a matrix of variables can be PSD while satisfying some equalities. One relaxes the

## 3 A few simple SOS preliminaries

A well-known basic fact (following from the Fundamental Theorem of Algebra) is that every nonnegative *univariate* polynomial is SOS:

**Fact 3.1.** *Suppose* $p \in \mathbb{R}[X_1]$ *is a univariate real polynomial such that* $p(t) \geq 0$ *for all real* $t$*. Then* $p$ *is SOS; i.e.,* $\vdash_{\deg(p)} p \geq 0$.

The following related result is credited in [38] to Fekete and Markov–Lukács, with reference also to [41]:

**Fact 3.2.** *Suppose* $p \in \mathbb{R}[Y]$ *is a univariate real polynomial of degree* $k$ *such that* $p(t) \geq 0$ *for all real* $a \leq t \leq b$.
*If* $k$ *is odd then*

$$Y \geq a,\ b \geq Y \quad \vdash_k \quad p \geq 0.$$

*If* $k$ *is even then*

$$(b - Y)(Y - a) \geq 0 \quad \vdash_k \quad p \geq 0.$$

We now give some additional simple SOS proofs:

**Fact 3.3.** $Y^2 \leq 1 \vdash_2 Y \leq 1, Y \geq -1$.

*Proof.* The first follows from $1 - Y = \frac{1}{2}(1 - Y)^2 + \frac{1}{2}(1 - Y^2)$. The second follows by replacing $Y$ by $-Y$. □

**Fact 3.4.** $\{Y \leq 1, Y \geq -1\} \vdash_3 Y^2 \leq 1$.

*Proof.* $1 - Y^2 = \frac{1}{2}(1 + Y)^2(1 - Y) + \frac{1}{2}(1 - Y)^2(1 + Y)$. □

We will need an SOS proof of the fact that $Y, Z \in \{-1, 1\} \Rightarrow \frac{Y-Z}{2} \in \{-1, 0, 1\}$:

**Fact 3.5.** $Y^2 = 1, Z^2 = 1 \vdash_3 (\frac{Y-Z}{2}) = (\frac{Y-Z}{2})^3$.

*Proof.* $(\frac{Y-Z}{2}) - (\frac{Y-Z}{2})^3 = (\frac{3}{8}Z - \frac{1}{8}Y)(Y^2 - 1) + (\frac{1}{8}Z - \frac{3}{8}Y)(Z^2 - 1)$. □

**Fact 3.6.** *Suppose that* $A \vdash_k Y \geq -1, Y \leq 1$ *and that* $B \vdash_\ell Z \geq W, Z \geq -W$*. Then* $A \cup B \vdash_{k+\ell} Z \geq YW$.

*Proof.* $Z - YW = \frac{1}{2}(Z - W)(1 + Y) + \frac{1}{2}(Z + W)(1 - Y)$. □

**Fact 3.7.** *Suppose that* $A \vdash_k Y' \geq Y$ *and* $B \vdash_\ell Z' \geq Z$*. Further suppose* $A' \vdash_{k'} Y' \geq 0$ *and* $B' \vdash_{\ell'} Z \geq 0$*. Then* $A \cup B \cup A' \cup B' \vdash_{\max\{k+\ell', k'+\ell\}} Y'Z' \geq YZ$.

*Proof.* This follows from $Y'Z' - YZ = Y'(Z' - Z) + Z(Y' - Y)$. □

We now move to Hölder-type inequalities.

**Fact 3.8.** $\vdash_2 YZ \leq \frac{1}{2}Y^2 + \frac{1}{2}Z^2$.

*Proof.* $\frac{1}{2}Y^2 + \frac{1}{2}Z^2 - YZ = \frac{1}{2}(Y - Z)^2$. □

More generally, by replacing $Y$ with $\epsilon^{1/2}Y$ and $Z$ with $\epsilon^{-1/2}Z$, we obtain:

equalities to two-sided inequalities with some small tolerance $\delta = 2^{-\mathrm{poly}(n)}$, allowing one to run Ellipsoid. If Ellipsoid returns a feasible solution it can be made truly PSD at the expense of adding slightly more slack in the equalities. By virtue of the compactness, this can adjusted to give a valid SOS proof of $-1 + \delta' \geq 0$.

**Fact 3.9.** $\vdash_2 YZ \le \frac{\epsilon}{2}Y^2 + \frac{1}{2\epsilon}Z^2$ *for any real* $\epsilon > 0$.

We would also like Young's inequality for conjugate Hölder exponents $(\frac{4}{3}, 4)$, but stating it needs a trick:

**Fact 3.10.** $\vdash_4 Y^3 Z \le \frac{3}{4}Y^4 + \frac{1}{4}Z^4$.

*Proof.* $\frac{3}{4}Y^4 + \frac{1}{4}Z^4 - Y^3 Z = (\frac{3}{4}Y^2 + \frac{1}{2}YZ + \frac{1}{4}Z^2)(Y-Z)^2$
$= (\frac{1}{2}Y^2 + \frac{1}{4}(Y+Z)^2)(Y-Z)^2 = \frac{1}{2}Y^2(Y-Z)^2 + \frac{1}{4}(Y^2 - Z^2)^2.$ □

By replacing $Y$ with $\epsilon^{1/4}Y$ and $Z$ with $\epsilon^{-3/4}Z$, we obtain:

**Fact 3.11.** $\vdash_4 Y^3 Z \le \frac{3\epsilon}{4}Y^4 + \frac{1}{4\epsilon^3}Z^4$ *for any real* $\epsilon > 0$.

**Fact 3.12.** *If* $A \vdash_k Y \ge 0$ *and* $A \vdash_k Y \le Z$*, then* $A \vdash_{2k} Y^2 \le Z^2$.

*Proof.* We can deduce $A \vdash_k Z \ge 0$ and therefore $A \vdash_k Z + Y \ge 0$ using Fact 2.2. The result now follows from Fact 2.4 applied to $Z^2 - Y^2 = (Z+Y)(Z-Y)$. □

**Fact 3.13.** $\vdash_2 \operatorname{avg}_{i\in[n]}[X_i^2] \ge (\operatorname{avg}_{i\in[n]}[X_i])^2$.

*Proof.* $\operatorname{avg}_{i\in[n]}[X_i^2] - (\operatorname{avg}_{i\in[n]}[X_i])^2 = \operatorname{avg}_{i,j\in[n]}[\frac{1}{2}(X_i - X_j)^2]$. □

# 4 SOS proofs of hypercontractivity

In the remainder of the work we will use some standard notions from analysis of Boolean functions; see, e.g., [46]. All of our main results will require SOS proofs of the well-known hypercontractivity theorems on $\{-1,1\}^n$, first proved by Bonami [7]. To state them, recall that any function $f : \{-1,1\}^n \to \mathbb{R}$ can be viewed as a multilinear polynomial,

$$f(x) = \sum_{S\subseteq[n]} \widehat{f}(S) \prod_{i\in S} x_i, \qquad \text{where } \widehat{f}(S) = \mathop{\mathbf{E}}_{\boldsymbol{x}\sim\{-1,1\}^n}[f(\boldsymbol{x}) \prod_{i\in S} \boldsymbol{x}_i]. \tag{9}$$

Then for $\rho \in \mathbb{R}$, the linear operator $\mathrm{T}_\rho$ is defined by mapping the above function to

$$\mathrm{T}_\rho f(x) = \sum_{S\subseteq[n]} \rho^{|S|} \widehat{f}(S) \prod_{i\in S} x_i.$$

Now the $p = 2$, $q \ge 2$ cases of hypercontractivity can be stated as follows:

**Theorem 4.1.** *Let* $f : \{-1,1\}^n \to \mathbb{R}$. *Then for any real* $q \ge 2$,

$$\mathop{\mathbf{E}}_{\boldsymbol{x}\sim\{-1,1\}^n}[|\mathrm{T}_{\frac{1}{\sqrt{q-1}}} f(\boldsymbol{x})|^q] \le \mathop{\mathbf{E}}_{\boldsymbol{x}\sim\{-1,1\}^n}[f(\boldsymbol{x})^2]^{q/2}.$$

**Theorem 4.2.** *Let* $f : \{-1,1\}^n \to \mathbb{R}$ *have degree at most* $k$. *Then for any real* $q \ge 2$,

$$\mathop{\mathbf{E}}_{\boldsymbol{x}\sim\{-1,1\}^n}[|f(\boldsymbol{x})|^q] \le (q-1)^{(q/2)k} \cdot \mathop{\mathbf{E}}_{\boldsymbol{x}\sim\{-1,1\}^n}[f(\boldsymbol{x})^2]^{q/2}.$$

Note that Theorem 4.2 follows immediately from Theorem 4.1 in case $f$ is homogeneous of degree $k$. It is also known that Theorem 4.1 and Theorem 4.2 (even its homogeneous version) are "equivalent", in the sense that one can be derived from the other using various analytic tricks.

As mentioned, we would ideally like to give SOS proofs of these theorems. In order to even *state* the theorems as polynomial inequalities it is required that $q$ be an even integer. For example, when $q = 4$ we may try to SOS-prove

$$\mathop{\mathbf{E}}_{\boldsymbol{x}\sim\{-1,1\}^n}[(\mathrm{T}_{\frac{1}{\sqrt{q-1}}} f(\boldsymbol{x}))^4] \leq \mathop{\mathbf{E}}_{\boldsymbol{x}\sim\{-1,1\}^n}[f(\boldsymbol{x})^2]^2.$$

The meaning of this is that *the $2^n$ Fourier coefficients of $f$ are the indeterminates*; i.e., we work over the ring $\mathbb{R}[\widehat{f}(\emptyset), \widehat{f}(\{1\}), \dots, \widehat{f}([n])]$ and would like to show that

$$\mathop{\mathbf{E}}_{\boldsymbol{x}\sim\{-1,1\}^n}[f(\boldsymbol{x})^2]^2 - \mathop{\mathbf{E}}_{\boldsymbol{x}\sim\{-1,1\}^n}[(\mathrm{T}_{\frac{1}{\sqrt{q-1}}} f(\boldsymbol{x}))^4]$$

is a sum of squares of polynomials over the indeterminates $\widehat{f}(S)$. Sometimes we will instead use the $2^n$ indeterminates "$f(x)$" for $x \in \{-1,1\}^n$ — note that this is completely equivalent because the $f(x)$'s are homogeneous linear forms in the $\widehat{f}(S)$'s and vice versa; see (9).

When $q$ is an even integer it is well known that Theorem 4.2 has a much simpler, "almost combinatorial" proof. For example, Bonami's original paper proved the homogeneous version of Theorem 4.2 for even integer $q$ using nothing more "analytic" than absolute values and Cauchy–Schwarz. (Her proof even obtains a slightly sharper constant than $(q-1)^{(q/2)k}$.) The inductive proof of Theorem 4.2 for $q = 4$ presented in [42] is simpler still, using only Cauchy–Schwarz. It is not hard to check that these remarks also apply to Theorem 4.1.

Nevertheless, it's not completely trivial to obtain SOS proofs of Theorems 4.1 and 4.2 when $q$ is an even integer, simply because the Cauchy–Schwarz inequality, $\mathbf{E}[fg] \leq \sqrt{\mathbf{E}[f^2]}\sqrt{\mathbf{E}[g^2]}$, has square-roots in it. The natural substitute is the inequality $\mathbf{E}[fg] \leq \frac{1}{2}\mathbf{E}[f^2] + \frac{1}{2}\mathbf{E}[g^2]$ (see Fact 3.8). However fitting this into the known proof of, say, the $q = 4$ case of Theorem 4.2 seems to require an extra trick: moving to a "two-function" version of the statement. This is what was done by Barak et al. in [3], wherein the following was shown:

**Theorem 4.3.** *(SOS proof of the two-function, $q = 4$ version of Theorem 4.2, [3].)*
*Let $n, k_1, k_2 \in \mathbb{N}$. For each $j = 1, 2$ and each $S \subseteq [n]$ of cardinality at most $k_j$, introduce an indeterminate $\widehat{f_j}(S)$. For $x \in \{-1,1\}^n$, let $f_j(x)$ denote $\sum_S \widehat{f_j}(S) \prod_{i\in S} x_i$. Then*

$$\vdash_4 \quad \mathop{\mathbf{E}}_{\boldsymbol{x}\sim\{-1,1\}^n}[f_1(\boldsymbol{x})^2 f_2(\boldsymbol{x})^2] \leq 3^{k_1+k_2} \cdot \mathop{\mathbf{E}}_{\boldsymbol{x}\sim\{-1,1\}^n}[f_1(\boldsymbol{x})^2] \cdot \mathop{\mathbf{E}}_{\boldsymbol{x}\sim\{-1,1\}^n}[f_2(\boldsymbol{x})^2].$$

Here we similarly give an SOS proof of the $q = 4$ case of Theorem 4.1. We will need a more general statement which allows for some of the $\pm 1$ random variables to be replaced by Gaussians; this idea is also from [42].

**Theorem 4.4.** *(SOS proof of the two-function, $q = 4$ version of Theorem 4.1.)*
*Let $n \in \mathbb{N}$. For each $j = 1, 2$ and each $S \subseteq [n]$, introduce an indeterminate $\widehat{f_j}(S)$. For each $z = (z_1, \dots, z_n) \in \mathbb{R}^n$, let*

$$f_j(z) = \sum_{S\subseteq[n]} \widehat{f_j}(S) \prod_{i\in S} z_i, \qquad \mathrm{T}_{\frac{1}{\sqrt{3}}} f_j(z) = \sum_{S\subseteq[n]} (\tfrac{1}{\sqrt{3}})^{|S|} \widehat{f_j}(S) \prod_{i\in S} z_i;$$

*these are homogeneous linear polynomials in the indeterminates. Let $\boldsymbol{z} = (\boldsymbol{z}_1, \dots, \boldsymbol{z}_n)$ be a random vector in which the components $\boldsymbol{z}_i$ are independent and satisfy $\mathbf{E}[\boldsymbol{z}_i] = \mathbf{E}[\boldsymbol{z}_i^3] = 0$, $\mathbf{E}[\boldsymbol{z}_i^2] = 1$, $\mathbf{E}[\boldsymbol{z}_i^4] \leq 9$. (For example, Rademachers and standard Gaussians qualify.) Then*

$$\vdash_4 \quad \mathop{\mathbf{E}}_{\boldsymbol{z}}[(\mathrm{T}_{\frac{1}{\sqrt{3}}} f_1(\boldsymbol{z}))^2 \cdot (\mathrm{T}_{\frac{1}{\sqrt{3}}} f_2(\boldsymbol{z}))^2] \leq \mathop{\mathbf{E}}_{\boldsymbol{z}}[f_1(\boldsymbol{z})^2] \cdot \mathop{\mathbf{E}}_{\boldsymbol{z}}[f_2(\boldsymbol{z})^2].$$

*In particular,*

$$\vdash_4 \quad \mathop{\mathbf{E}}_{\boldsymbol{z}}[(\mathrm{T}_{\frac{1}{\sqrt{3}}} f_1(\boldsymbol{z}))^4] \leq \mathop{\mathbf{E}}_{\boldsymbol{z}}[f_1(\boldsymbol{z})^2]^2.$$

*Proof.* The proof of the theorem is by induction on $n$. For $n = 0$ we need to show $\vdash_4 \widehat{f}_1(\emptyset)^2\widehat{f}_2(\emptyset)^2 \le \widehat{f}_1(\emptyset)^2\widehat{f}_2(\emptyset)^2$, which is trivial. For general $n \ge 1$ and $(z_1, \ldots, z_n) \in \mathbb{R}^n$ we can express $f_j(z_1, \ldots, z_n) = z_n d_j(z') + e_j(z')$, where $z' \in \mathbb{R}^{n-1}$ denotes $(z_1, \ldots, z_{n-1})$,

$$d_j(z') = \sum_{S \ni n} \widehat{f}_j(S) \prod_{i \in S \setminus \{n\}} z_i,$$

$$e_j(z') = \sum_{S \not\ni n} \widehat{f}_j(S) \prod_{i \in S} z_i.$$

Now

$$\begin{aligned}
&\mathop{\mathbf{E}}_{\boldsymbol{x} \sim \{-1,1\}^n} [(\mathrm{T}_{\frac{1}{\sqrt{3}}} f_1(\boldsymbol{z}))^2 \cdot (\mathrm{T}_{\frac{1}{\sqrt{3}}} f_2(\boldsymbol{z}))^2] \\
&= \mathop{\mathbf{E}}_{\boldsymbol{z}} \left[ \left( \tfrac{1}{\sqrt{3}} \boldsymbol{z}_n \cdot \mathrm{T}_{\frac{1}{\sqrt{3}}} d_1(\boldsymbol{z}') + \mathrm{T}_{\frac{1}{\sqrt{3}}} e_1(\boldsymbol{z}') \right)^2 \left( \tfrac{1}{\sqrt{3}} \boldsymbol{z}_n \cdot \mathrm{T}_{\frac{1}{\sqrt{3}}} d_2(\boldsymbol{z}') + \mathrm{T}_{\frac{1}{\sqrt{3}}} e_2(\boldsymbol{z}') \right)^2 \right] \\
&= \mathop{\mathbf{E}}_{\boldsymbol{z}} \left[ \left( \tfrac{1}{3} \boldsymbol{z}_n^2 (\mathrm{T}d_1)^2 + \tfrac{2}{\sqrt{3}} \boldsymbol{z}_n (\mathrm{T}d_1)(\mathrm{T}e_1) + (\mathrm{T}e_1)^2 \right) \left( \tfrac{1}{3} \boldsymbol{z}_n^2 (\mathrm{T}d_2)^2 + \tfrac{2}{\sqrt{3}} \boldsymbol{z}_n (\mathrm{T}d_2)(\mathrm{T}e_2) + (\mathrm{T}e_2)^2 \right) \right],
\end{aligned}$$

where we introduced the shorthand $(\mathrm{T}d_j)$ for $\mathrm{T}_{\frac{1}{\sqrt{3}}} d_j(\boldsymbol{z}')$ (and similarly for $e_j$). We continue by expanding the product and using linearity of expectation, $\mathbf{E}[\boldsymbol{z}_n] = \mathbf{E}[\boldsymbol{z}_n^3] = 0$, $\mathbf{E}[\boldsymbol{z}_n^2] = 1$; thus the above equals

$$\begin{aligned}
\tfrac{1}{9}\mathbf{E}[\boldsymbol{z}_n^4] \mathop{\mathbf{E}}_{\boldsymbol{z}'}[(\mathrm{T}d_1)^2(\mathrm{T}d_2)^2] + \tfrac{1}{3}\mathop{\mathbf{E}}_{\boldsymbol{z}'}[(\mathrm{T}d_1)^2(\mathrm{T}e_2)^2] + \tfrac{1}{3}\mathop{\mathbf{E}}_{\boldsymbol{z}'}[(\mathrm{T}d_2)^2(\mathrm{T}e_1)^2] + \mathop{\mathbf{E}}_{\boldsymbol{z}'}[(\mathrm{T}e_1)^2(\mathrm{T}e_2)^2] \\
+ \tfrac{4}{3}\mathop{\mathbf{E}}_{\boldsymbol{z}'}[(\mathrm{T}d_1)(\mathrm{T}e_2) \cdot (\mathrm{T}d_2)(\mathrm{T}e_1)].
\end{aligned}$$

Using Fact 3.8 we have

$$\vdash_4 \quad \tfrac{4}{3}\mathop{\mathbf{E}}_{\boldsymbol{z}'}[(\mathrm{T}d_1)(\mathrm{T}e_2) \cdot (\mathrm{T}d_2)(\mathrm{T}e_1)] \le \tfrac{2}{3}\mathop{\mathbf{E}}_{\boldsymbol{z}'}[(\mathrm{T}d_1)^2(\mathrm{T}e_2)^2] + \tfrac{2}{3}\mathop{\mathbf{E}}_{\boldsymbol{z}'}[(\mathrm{T}d_2)^2(\mathrm{T}e_1)^2].$$

By our assumption $\mathbf{E}[\boldsymbol{z}_n^2] \le 9$ we have $\vdash_4 \frac{1}{9}\mathbf{E}[\boldsymbol{z}_n^4]\,\mathbf{E}_{\boldsymbol{z}'}[(\mathrm{T}d_1)^2(\mathrm{T}d_2)^2] \le \mathbf{E}_{\boldsymbol{z}'}[(\mathrm{T}d_1)^2(\mathrm{T}d_2)^2]$; here we are using the fact that $\mathbf{E}_{\boldsymbol{z}'}[(\mathrm{T}d_1)^2(\mathrm{T}d_2)^2]$ is SOS.[3] Thus we have shown

$$\begin{aligned}
\vdash_4 \quad &\mathop{\mathbf{E}}_{\boldsymbol{z}}[(\mathrm{T}_{\frac{1}{\sqrt{3}}} f_1(\boldsymbol{z}))^2 \cdot (\mathrm{T}_{\frac{1}{\sqrt{3}}} f_2(\boldsymbol{z}))^2] \\
&\le \mathop{\mathbf{E}}_{\boldsymbol{z}'}[(\mathrm{T}d_1)^2(\mathrm{T}d_2)^2] + \mathop{\mathbf{E}}_{\boldsymbol{z}'}[(\mathrm{T}d_1)^2(\mathrm{T}e_2)^2] + \mathop{\mathbf{E}}_{\boldsymbol{z}'}[(\mathrm{T}d_2)^2(\mathrm{T}e_1)^2] + \mathop{\mathbf{E}}_{\boldsymbol{z}'}[(\mathrm{T}e_1)^2(\mathrm{T}d_e)^2].
\end{aligned}$$

We use induction on each of the four terms above and deduce

$$\begin{aligned}
\vdash_4 \quad &\mathop{\mathbf{E}}_{\boldsymbol{z}}[(\mathrm{T}_{\frac{1}{\sqrt{3}}} f_1(\boldsymbol{z}))^2 \cdot (\mathrm{T}_{\frac{1}{\sqrt{3}}} f_2(\boldsymbol{z}))^2] \\
&\le \mathop{\mathbf{E}}_{\boldsymbol{z}'}[d_1(\boldsymbol{z}')^2] \mathop{\mathbf{E}}_{\boldsymbol{z}'}[d_2(\boldsymbol{z}')^2] + \mathop{\mathbf{E}}_{\boldsymbol{z}'}[d_1(\boldsymbol{z}')^2] \mathop{\mathbf{E}}_{\boldsymbol{z}'}[e_2(\boldsymbol{z}')^2] + \mathop{\mathbf{E}}_{\boldsymbol{z}'}[d_2(\boldsymbol{z}')^2] \mathop{\mathbf{E}}_{\boldsymbol{z}'}[e_1(\boldsymbol{z}')^2] + \mathop{\mathbf{E}}_{\boldsymbol{z}'}[e_1(\boldsymbol{z}')^2] \mathop{\mathbf{E}}_{\boldsymbol{z}'}[e_2(\boldsymbol{z}')^2] \\
&= \mathop{\mathbf{E}}_{\boldsymbol{z}'}[d_1(\boldsymbol{z}')^2 + e_1(\boldsymbol{z}')^2] \cdot \mathop{\mathbf{E}}_{\boldsymbol{z}'}[d_1(\boldsymbol{z}')^2 + e_1(\boldsymbol{z}')^2]
\end{aligned}$$

But it is easily verified that $\mathbf{E}_{\boldsymbol{z}}[f_j(\boldsymbol{z})^2] = \mathbf{E}_{\boldsymbol{z}'}[d_j(\boldsymbol{z}')^2 + e_j(\boldsymbol{z}')^2]$, completing the induction. □

[3] When $\boldsymbol{z}'$ is a discrete random vector this is obvious. In the general case, note that the coefficients of the polynomial in question are finite mixed moments of $\boldsymbol{z}'$. By Carathéodory's convex hull theorem we can match any finite number of moments of $\boldsymbol{z}'$ using some discrete random vector $\boldsymbol{z}''$, thereby reducing SOS-verification to the discrete case. We will use this observation in the sequel without additional comment.

From this we can deduce Theorem 4.3 with the more general class of random variables. (Alternately, it is not hard to obtain the following by generalizing the proof in [3].)

**Corollary 4.5.** *Theorem 4.3 also holds with the more general type of random vector* $\boldsymbol{z}$ *from Theorem 4.4 in place of* $\boldsymbol{x} \sim \{-1,1\}^n$.

*Proof.* Begin by defining

$$\widehat{g_j}(S) = \begin{cases} 0 & \text{if } |S| > k_j, \\ \sqrt{3}^{|S|}\widehat{f_j}(S) & \text{if } |S| \le k_j \end{cases}$$

for $j = 1, 2$, and then applying Theorem 4.4 to $g_1$, $g_2$. This yields

$$\vdash_4 \quad \mathop{\mathbf{E}}_{\boldsymbol{z}}[f_1(\boldsymbol{z})^2 f_2(\boldsymbol{z})^2] \le \mathop{\mathbf{E}}_{\boldsymbol{z}}[\mathrm{T}_{\sqrt{3}} f_1(\boldsymbol{z})^2] \cdot \mathop{\mathbf{E}}_{\boldsymbol{z}}[\mathrm{T}_{\sqrt{3}} f_2(\boldsymbol{z})^2].$$

By a standard computation we have

$$\mathop{\mathbf{E}}_{\boldsymbol{z}}[\mathrm{T}_{\sqrt{3}} f_j(\boldsymbol{z})^2] = \sum_{i=0}^{k_j} 3^i \cdot W_j^{=i}, \quad \text{where} \quad W_j^{=i} = \sum_{|S|=i} \widehat{f_j}(S)^2, \quad j = 1, 2.$$

We also have $\mathbf{E}_{\boldsymbol{z}}[f_j(\boldsymbol{z})^2] = \sum_{i=0}^{k_j} W_j^{=i}$. Thus to complete the proof it remains to show

$$\vdash_4 \quad \left(\sum_{i=0}^{k_1} 3^i \cdot W_1^{=i}\right)\left(\sum_{i'=0}^{k_2} 3^{i'} \cdot W_2^{=i'}\right) \le \left(\sum_{i=0}^{k_1} 3^{k_1} \cdot W_1^{=i}\right)\left(\sum_{i'=0}^{k_2} 3^{k_2} \cdot W_2^{=i'}\right).$$

But after distributing out both products, this is immediate from

$$\vdash_4 \quad 3^{i+i'} \cdot W_1^{=i} \cdot W_2^{=i'} \le 3^{k_1+k_2} \cdot W_1^{=i} \cdot W_2^{=i'}$$

for each $0 \le i \le k_1, 0 \le i' \le k_2$. ☐

We would also like to have an SOS proof of Theorem 4.2 for even integers $q > 4$. We content ourselves with the following slightly weaker result, the proof of which follows easily from Corollary 4.5:

**Theorem 4.6.** *(SOS proof of a weakened version of the two-function, even integer* $q$ *case of Theorem 4.2.)*

*Let* $n, r, k_1, k_2, \dots, k_{2^r} \in \mathbb{N}$. *For each* $j \in [2^r]$ *and each* $S \subseteq [n]$ *of cardinality at most* $k_j$, *introduce an indeterminate* $\widehat{f_j}(S)$. *Let* $f_1(z), \dots, f_{2^r}(z)$ *and random vector* $\boldsymbol{z}$ *be as in Theorem 4.4. Then*

$$\vdash_{2^{r+1}} \quad \mathop{\mathbf{E}}_{\boldsymbol{z}}\left[\prod_{j=1}^{2^r} f_j(\boldsymbol{z})^2\right] \le 3^{r(k_1+\cdots+k_{2^r})} \cdot \prod_{j=1}^{2^r} \mathop{\mathbf{E}}_{\boldsymbol{z}}[f_j(\boldsymbol{z})^2].$$

*Proof.* The proof is by induction on $r$. The $r = 0$ case is trivial. For $r \ge 1$, define

$$F_1(z) = \prod_{j=1}^{2^{r-1}} f_j(z), \qquad F_2(z) = \prod_{j=2^{r-1}+1}^{2^r} f_j(z).$$

Note these are degree-$2^{r-1}$ in the indeterminates. Further, one may express

$$F_1(z) = \sum_{\substack{T \subseteq [n] \\ |T| \le k_1 + \cdots + k_{2^{r-1}}}} \widehat{f}(T) \prod_{i \in T} z_i,$$

where $\widehat{f}(T)$ denotes a degree-$2^{r-1}$ polynomial in the indeterminates, and similarly for $F_2$. Thus we may apply Corollary 4.5 to $F_1$ and $F_2$ and deduce

$$\vdash_{2^{r+1}} \quad \mathop{\mathbf{E}}_{\boldsymbol{z}}\left[\prod_{j=1}^{2^r} f_j(\boldsymbol{z})^2\right] \leq 3^{k_1+\cdots+k_{2^r}} \cdot \mathop{\mathbf{E}}_{\boldsymbol{z}}\left[\prod_{j=1}^{2^{r-1}} f_j(\boldsymbol{z})^2\right] \cdot \mathop{\mathbf{E}}_{\boldsymbol{z}}\left[\prod_{j=2^{r-1}+1}^{2^r} f_j(\boldsymbol{z})^2\right]. \tag{10}$$

By induction we have

$$\vdash_{2^r} \quad \mathop{\mathbf{E}}_{\boldsymbol{z}}\left[\prod_{j=1}^{2^{r-1}} f_j(\boldsymbol{z})^2\right] \leq 3^{(r-1)(k_1+\cdots+k_{2^{r-1}})} \cdot \prod_{j=1}^{2^{r-1}} \mathop{\mathbf{E}}_{\boldsymbol{z}}[f_j(\boldsymbol{z})^2],$$

$$\vdash_{2^r} \quad \mathop{\mathbf{E}}_{\boldsymbol{z}}\left[\prod_{j=2^{r-1}+1}^{2^r} f_j(\boldsymbol{z})^2\right] \leq 3^{(r-1)(k_{2^{r-1}+1}+\cdots+k_{2^r})} \cdot \prod_{j=2^{r-1}+1}^{2^r} \mathop{\mathbf{E}}_{\boldsymbol{z}}[f_j(\boldsymbol{z})^2],$$

and all four expressions above are SOS of degree $2^r$. Combining these via Fact 3.7 yields

$$\vdash_{2^{r+1}} \quad \mathop{\mathbf{E}}_{\boldsymbol{z}}\left[\prod_{j=1}^{2^{r-1}} f_j(x)^2\right] \cdot \mathop{\mathbf{E}}_{\boldsymbol{z}}\left[\prod_{j=2^{r-1}+1}^{2^r} f_j(x)^2\right] \leq 3^{(r-1)(k_1+\cdots+k_{2^r})} \cdot \prod_{j=1}^{2^{r-1}} \mathop{\mathbf{E}}_{\boldsymbol{z}}[f_j(\boldsymbol{z})^2] \cdot \prod_{j=2^{r-1}+1}^{2^r} \mathop{\mathbf{E}}_{\boldsymbol{z}}[f_j(\boldsymbol{z})^2],$$

which taken together with (10) completes the induction. ☐

**Corollary 4.7.** *(SOS proof of a weakened version of the even integer $q$ case of Theorem 4.2.)*

*Let $n, k \in \mathbb{N}$. For each $S \subseteq [n]$ of cardinality at most $k$, introduce an indeterminate $\widehat{f}(S)$. Let $f(z)$ and random vector $\boldsymbol{z}$ be as in Theorem 4.4. Then for any even integer $q \geq 2$,*

$$\vdash_{2^{\lceil \log_2 q \rceil}} \quad \mathop{\mathbf{E}}_{\boldsymbol{z}}[f(\boldsymbol{z})^q] \leq \sqrt{3}^{(q\lceil \log_2 q \rceil - q)k} \cdot \mathop{\mathbf{E}}_{\boldsymbol{z}}[f_j(\boldsymbol{z})^2]^{q/2}.$$

*Proof.* Take $r = \lceil \log_2 q \rceil - 1$, $f_1 = \cdots = f_{q/2} = f$, $f_{q/2+1} = \cdots = f_{2^r} = 1$ in Theorem 4.6. ☐

# 5 SOS proofs of SSE in the Noisy Hypercube, and KKL

## 5.1 An SOS proof of small-set expansion in the noisy hypercube

The following well-known theorem concerning small-set expansion (SSE) in the hypercube is due to Kahn, Kalai, and Linial [23]:

**Noisy Hypercube SSE Theorem.** *Let $f : \{-1,1\}^n \to \{-1,0,1\}$. Then for any $0 \leq \rho \leq 1$,*

$$\mathbf{Stab}_\rho[f] \leq \mathbf{E}[f^2]^{2/(1+\rho)},$$

*where $\mathbf{Stab}_\rho[f]$ denotes $\langle f, \mathrm{T}_\rho f\rangle = \|\mathrm{T}_{\sqrt{\rho}} f\|_2^2$.*

*Proof.*

$$\mathbf{Stab}_\rho[f] = \|\mathrm{T}_{\sqrt{\rho}} f\|_2^2 \leq \|f\|_{1+\rho}^2 = \mathbf{E}[|f|^{1+\rho}]^{2/(1+\rho)} = \mathbf{E}[f^2]^{2/(1+\rho)},$$

where the inequality is hypercontractivity (the Hölder dual of Theorem 4.1). ☐

We remark on two special cases:

$$\mathbf{Stab}_{\frac{1}{3}}[f] \leq \mathbf{E}[f^2]^{3/2}, \qquad \mathbf{Stab}_{\frac{1}{\sqrt{3}}}[f] \leq \mathbf{E}[f^2]^{3-\sqrt{3}} \leq \mathbf{E}[f^2]^{1.2679}.$$

We do not know how to obtain a low-degree SOS proof of either inequality. Nevertheless, we come close in the following theorem. We remark again that its proofs bears some similarities to the Barak et al. analysis of the KV Unique-games instances [3].

**Theorem 5.1.** *(SOS proof of a weakened special case of the Noisy Hypercube SSE Theorem.)*
*Let $n \in \mathbb{N}$, and for each $x \in \{-1,1\}^n$ let $f(x)$ be an indeterminate. Then for any real $\epsilon > 0$,*

$$\{f(x) = f(x)^3 : \forall x\} \quad \vdash_4 \quad \mathbf{Stab}_{\frac{1}{\sqrt{3}}}[f] \leq \mathop{\mathbf{E}}_{\boldsymbol{x}}[f(\boldsymbol{x})^2]\left(\tfrac{3\epsilon}{4} + \tfrac{1}{4\epsilon^3}\mathbf{E}[f(\boldsymbol{x})^2]\right).$$

**Remark 5.2.** From this we can deduce that if $f : \{-1,1\}^n \to \{-1,0,1\}$ is an ordinary function then $\mathbf{Stab}_{\frac{1}{\sqrt{3}}}[f] \leq \mathbf{E}[f(\boldsymbol{x})^2]^{5/4}$, by taking $\epsilon = \mathbf{E}[f(\boldsymbol{x})^2]^{1/4}$.

*Proof.* From Fact 3.11 (and the trivial fact $Y = Y^3 \vdash_4 Yp^2 = Y^4$) we may easily deduce

$$Y = Y^3 \quad \vdash_4 \quad YZ \leq \tfrac{3\epsilon}{4}Y^2 + \tfrac{1}{4\epsilon^3}Z^4.$$

Since $\mathbf{Stab}_{\frac{1}{\sqrt{3}}}[f] = \mathbf{E}_{\boldsymbol{x}}[f(\boldsymbol{x})\mathrm{T}_{\frac{1}{\sqrt{3}}}f(\boldsymbol{x})]$ we may therefore obtain

$$\{f(x) = f(x)^3 : \forall x\} \quad \vdash_4 \quad \mathbf{Stab}_{\frac{1}{\sqrt{3}}}[f] \leq \tfrac{3\epsilon}{4}\mathop{\mathbf{E}}_{\boldsymbol{x}}[f(\boldsymbol{x})^2] + \tfrac{1}{4\epsilon^3}\mathop{\mathbf{E}}_{\boldsymbol{x}}[\mathrm{T}_{\frac{1}{\sqrt{3}}}f(\boldsymbol{x})^4].$$

The result now follows from Theorem 4.4. ☐

## 5.2 The KKL Theorem

With the Noisy Hypercube SSE Theorem in hand, we can now give an SOS proof of the famed KKL Theorem [23], the key ingredient in the analysis of the DKSV Balanced-Separator instances.

**Theorem 5.3.** *(SOS proof of the KKL Theorem.)*
*Let $n \in \mathbb{N}$, and for each $x \in \{-1,1\}^n$ let $f(x)$ be an indeterminate. Let $\tau$ be an indeterminate. Then for any reals $\epsilon > 0$, $K \geq 2$,*

$$\{f(x)^2 = 1 : \forall x\} \cup \{\mathbf{Inf}_i[f] \leq \tau : \forall i \in [n]\} \quad \vdash_4 \quad \mathbf{Var}[f] \leq \left(\frac{\sqrt{3}^{K-1}}{K}\left(\tfrac{3\epsilon}{4} + \tfrac{\tau}{4\epsilon^3}\right) + \tfrac{1}{K}\right)\mathbf{I}[f].$$

**Remark 5.4.** From this we can deduce that if $f : \{-1,1\}^n \to \{-1,1\}$ is an ordinary function and $\tau \leq \frac{1}{9}$ is a positive real such that $\mathbf{Inf}_i[f] \leq \tau$ for all $i$, then $\mathbf{I}[f] \geq \frac{1}{2}\log_9(\frac{9}{\tau}) \cdot \mathbf{Var}[f]$. This follows by taking $\epsilon = \tau^{1/4}$ and $K = \log_9(\frac{9}{\tau})$.

*Proof.* We may apply Theorem 5.1 to each of the derivative "functions"

$$\mathrm{D}_i f(x) = \frac{f(x^{(i\mapsto 1)}) - f(x^{(i\mapsto -1)})}{2}.$$

(These are actually sets of indeterminates, each of which is a homogeneous linear form in the indeterminates $f(x)$.) We can obtain the hypothesis $\mathrm{D}_i f(x) = \mathrm{D}_i f(x)^3$ from the hypotheses $f(x)^2 = 1$ via Fact 3.5. We deduce

$$\begin{aligned}\{f(x)^2 = 1 : \forall x\} \quad \vdash_4 \quad & \mathbf{Stab}_{\frac{1}{\sqrt{3}}}[\mathrm{D}_i f] \leq \mathop{\mathbf{E}}_{\boldsymbol{x}}[\mathrm{D}_i f(\boldsymbol{x})^2]\left(\tfrac{3\epsilon}{4} + \tfrac{1}{4\epsilon^3}\mathbf{E}[\mathrm{D}_i f(\boldsymbol{x})^2]\right) \\ \Leftrightarrow \quad & \sum_{S \ni i}(\tfrac{1}{\sqrt{3}})^{|S|-1}\widehat{f}(S)^2 \leq \mathbf{Inf}_i[f]\left(\tfrac{3\epsilon}{4} + \tfrac{1}{4\epsilon^3}\mathbf{Inf}_i[f]\right)\end{aligned}$$

for each $i \in [n]$. Further, since $\mathbf{Inf}_i[f]$ is SOS and of degree 2 we have

$$\mathbf{Inf}_i[f] \le \tau \quad \vdash_4 \quad \mathbf{Inf}_i[f] \cdot (\tfrac{\tau}{4\epsilon^3} - \tfrac{1}{4\epsilon^3}\mathbf{Inf}_i[f]) \ge 0.$$

Adding the previous two deductions yields

$$\{f(x)^2 = 1 : \forall x\} \cup \{\mathbf{Inf}_i[f] \le \tau : \forall i \in [n]\} \quad \vdash_4 \quad \sum_{S \ni i} (\tfrac{1}{\sqrt{3}})^{|S|-1} \widehat{f}(S)^2 \le \mathbf{Inf}_i[f] \cdot \big(\tfrac{3\epsilon}{4} + \tfrac{\tau}{4\epsilon^3}\big)$$

for each $i$. Now adding over all $i \in [n]$ gives

$$\{f(x)^2 = 1 : \forall x\} \cup \{\mathbf{Inf}_i[f] \le \tau : \forall i \in [n]\} \quad \vdash_4 \quad \sum_{S \subseteq [n]} |S|(\tfrac{1}{\sqrt{3}})^{|S|-1} \widehat{f}(S)^2 \le \big(\tfrac{3\epsilon}{4} + \tfrac{\tau}{4\epsilon^3}\big)\, \mathbf{I}[f].$$

Moreover, since $s(\frac{1}{\sqrt{3}})^{s-1} \ge K(\frac{1}{\sqrt{3}})^{K-1} - s(\frac{1}{\sqrt{3}})^{K-1}$ holds for all $s \in [n]$ (consider $s \le K$ and $s \ge K$), it follows that

$$\vdash_2 \quad \sum_{S \subseteq [n]} |S|(\tfrac{1}{\sqrt{3}})^{|S|-1} \widehat{f}(S)^2 \ge \tfrac{K}{\sqrt{3}^{K-1}}\, \mathbf{Var}[f] - \tfrac{1}{\sqrt{3}^{K-1}} \mathbf{I}[f].$$

By combining the previous two deductions and doing some rearranging, we obtain

$$\{f(x)^2 = 1 : \forall x\} \cup \{\mathbf{Inf}_i[f] \le \tau : \forall i \in [n]\} \quad \vdash_4 \quad \mathbf{Var}[f] \le \Big( \tfrac{\sqrt{3}^{K-1}}{K} \big(\tfrac{3\epsilon}{4} + \tfrac{\tau}{4\epsilon^3}\big) + \tfrac{1}{K} \Big)\, \mathbf{I}[f],$$

as claimed. □

We can now easily deduce (an SOS proof of) the fact that if $f : \{-1,1\}^n \to \{-1,1\}$ has constant variance and all its influences equal then its total influence is $\Omega(\log n)$. For the application to Balanced-Separator, we will in fact need a slightly more technical statement:

**Corollary 5.5.** *(SOS proof of KKL for equal-influence functions.)*
*Let $n \ge 81$ be an integer and for each $x \in \{-1,1\}^n$ let $f(x)$ be an indeterminate. Define*

$$A = \{f(x)^2 = 1 : \forall x\} \cup \{\mathbf{Inf}_i[f] \le \tau : \forall i \in [n]\} \\ \cup \{\mathbf{Inf}_i[f] = \mathbf{Inf}_j[f] : \forall i,j \in [n]\} \cup \{\mathbf{Var}[f] \ge \tfrac{3}{4}\} \cup \{\mathbf{I}[f] \le \tfrac{1}{20} \ln n\}.$$

*Then $A \vdash_4 -1 \ge 0$.*
*In fact, the result holds even if we change the equal-influences assumption $\{\mathbf{Inf}_i[f] = \mathbf{Inf}_j[f] : \forall i,j \in [n]\}$ to the weaker pair of assumptions $\{\mathbf{Inf}_i[f] = \mathbf{Inf}_j[f] : \forall i,j \le n/2\}$ and $\{\mathbf{Inf}_{i'}[f] = \mathbf{Inf}_{j'}[f] : \forall i',j' > n/2\}$ (assume $n$ even).*

*Proof.* We will prove the "in fact" statement, assuming $n$ is even. (The reader will see why the original statement is also true when $n$ is odd.) Define $\mathbf{I}^{(1)}[f] = \sum_{i \le n/2} \mathbf{Inf}_i[f]$ and $\mathbf{I}^{(2)}[f] = \sum_{i > n/2} \mathbf{Inf}_i[f]$, so $\mathbf{I}[f] = \mathbf{I}^{(1)}[f] + \mathbf{I}^{(2)}[f]$. Note that

$$\{\mathbf{Inf}_i[f] = \mathbf{Inf}_j[f] : \forall i,j \le n/2\} \quad \vdash_2 \quad \mathbf{Inf}_i[f] = \tfrac{2}{n}\mathbf{I}^{(1)}[f]$$

for each $i \le n/2$, and similarly for $i > n/2$. Since $\mathbf{I}^{(1)}[f], \mathbf{I}^{(2)}[f]$ are themselves SOS and of degree 2, we get

$$\{\mathbf{Inf}_i[f] = \mathbf{Inf}_j[f] : \forall i,j \le n/2 \,\&\, \forall i,j > n/2\} \quad \vdash_2 \quad \mathbf{Inf}_i[f] \le \tfrac{2}{n}\mathbf{I}^{(1)}[f] + \tfrac{2}{n}\mathbf{I}^{(2)}[f] = \tfrac{2}{n}\mathbf{I}[f]$$

for each $i \in [n]$. (Note that with the basic equal-influences assumption we can obtain the even stronger conclusion $\mathbf{Inf}_i[f] = \frac{1}{n}\mathbf{I}[f]$ for each $i \in [n]$.) We can now employ Theorem 5.3, replacing $\tau$ by $\frac{2}{n}\mathbf{I}[f]$. Using also $\mathbf{Var}[f] \geq \frac{3}{4}$, we obtain that for any reals $\epsilon > 0$, $K \geq 2$,

$$A \setminus \{\mathbf{I}[f] \leq \tfrac{1}{20}\ln n\} \quad \vdash_4 \quad \tfrac{3}{4} \leq \left(\frac{\sqrt{3}^{K-1}}{K}\left(\tfrac{3\epsilon}{4} + \tfrac{1}{2\epsilon^3 n}\mathbf{I}[f]\right) + \tfrac{1}{K}\right)\mathbf{I}[f].$$

Select $K = \log_9(9n^{1/2})$ and $\epsilon = n^{-1/8}$ to obtain

$$\begin{aligned} A \setminus \{\mathbf{I}[f] \leq \tfrac{1}{20}\ln n\} \quad \vdash_4 \quad \tfrac{3}{4} &\leq \left(\frac{n^{1/8}}{\log_9(9n^{1/2})}\left(\tfrac{3}{4}n^{-1/8} + \tfrac{1}{2}n^{-5/8}\mathbf{I}[f]\right) + \frac{1}{\log_9(9n^{1/2})}\right)\mathbf{I}[f] \\ &= \frac{7}{2\log_9(81n)}\mathbf{I}[f] + \frac{1}{n^{1/2}\log_9(81n)}\mathbf{I}[f]^2. \end{aligned} \tag{11}$$

We now employ $\mathbf{I}[f] \leq \frac{1}{20}\ln n$. Since $\mathbf{I}[f]$ is SOS and of degree 2 we also have

$$\mathbf{I}[f] \leq \tfrac{1}{20}\ln n \quad \vdash_4 \quad \mathbf{I}[f]^2 \leq \tfrac{1}{400}\ln^2 n.$$

Substituting this into (11) yields

$$A \quad \vdash_4 \quad \tfrac{3}{4} \leq \frac{7}{2\log_9(81n)} \cdot \tfrac{1}{20}\ln n + \frac{1}{n^{1/2}\log_9(81n)} \cdot \tfrac{1}{400}\ln^2 n \leq \tfrac{7}{20}\ln(3) \leq 0.4,$$

whence $A \vdash_4 -1 \geq 0$. $\square$

# 6 Analysis of the DKSV Balanced-Separator instances

We recall the Balanced-Separator problem: Given is an undirected multigraph $G = (V, E)$. It is required to find a cut $S \subseteq V$ with $\frac{1}{3} \leq \frac{|S|}{|V|} \leq \frac{2}{3}$ so as to minimize $\frac{E(S,\overline{S})}{|E|}$. The natural polynomial optimization formulation has an indeterminate $f(x)$ for each vertex $x \in V$:

$$\begin{aligned} \min \quad & \frac{1}{|E|}\sum_{(x,y)\in E}\left(\frac{f(x)-f(y)}{2}\right)^2 \\ \text{s.t.} \quad & f(x)^2 = 1 \quad \forall x \in V, \\ & \left(\frac{1}{|V|}\sum_{x\in V} f(x)\right)^2 \leq \frac{1}{9}. \end{aligned}$$

Thus as discussed in Section 2, the degree-4 SOS SDP hierarchy will use binary search to compute the largest $\beta$ for which

$$\{f(x)^2 = 1 : \forall x \in V\} \cup \left\{\left(\frac{1}{|V|}\sum_{x\in V} f(x)\right)^2 \leq \frac{1}{9}\right\} \cup \left\{\frac{1}{|E|}\sum_{(x,y)\in E}\left(\frac{f(x)-f(y)}{2}\right)^2 \leq \beta\right\} \quad \vdash_4 \quad -1 \geq 0.$$

**The DKSV instances.** We now recall the DKSV Balanced-Separator instances [12]. The instances $G = (V, E) = (V_N, E_N)$ are parameterized by primes $N$. Let $\mathcal{F} = \{-1,1\}^N \times \{-1,1\}^N$, thought of as the $2N$-dimensional hypercube graph. Let $\sigma$ act on elements $(x, y) \in \mathcal{F}$ by cyclic rotation of both halves: $\sigma(x, y) = (x_N, x_1, \ldots, x_{N-1}, y_N, y_1, \ldots, y_{N-1})$. The elements $\sigma, \sigma^2, \ldots, \sigma^N = id$ form a group acting on $\mathcal{F}$, partitioning it into orbits $\mathcal{O}_1, \ldots, \mathcal{O}_m$; 4 of these orbits have cardinality 1 and the remaining $(2^{2N}-4)/N$ have cardinality $N$. A cardinality-$N$ orbit $\mathcal{O}$ is called "nearly orthogonal" if for all distinct $(x, y), (x', y') \in \mathcal{O}$ it holds that $|\langle(x, y), (x', y')\rangle| \leq 8\sqrt{N \log N}$. Presuming that $N$ is sufficiently large, [12] shows that the number $n$ of nearly orthogonal orbits satisfies $(1-4/N^2)m \le n \le m$. (This implies $N = \Theta(\log n)$.) For typographic simplicity the nearly orthogonal orbits are assumed to be $\{\mathcal{O}_1, \dots, \mathcal{O}_n\}$, and this set is taken to be the vertex set $V$. We write $L \subseteq \mathcal{F}$ for the "leftover" elements contained in orbits $\mathcal{O}_{n+1}, \dots, \mathcal{O}_m$; writing $\epsilon = \frac{|L|}{2^{2N}}$ we have $\epsilon = O(1/N^2)$. The edges $E$ in $G$ are given by the usual hypercube edges in $\mathcal{F}$. More precisely, any pair $\mathcal{O}, \mathcal{O}' \in V$ have either $N$ or $0$ edges between them, according to whether or not there exist $(x,y) \in \mathcal{O}, (x',y') \in \mathcal{O}'$ at Hamming distance 1 in $\mathcal{F}$. There are no self-loops in $G$ because of the near orthogonality property. The set of edges $E$ is in one-to-one correspondence with a subset of (almost all the) hypercube edges in $\mathcal{F}$; specifically, all those not incident on $L$. The authors of [12] use the KKL Theorem to prove:

**Theorem 6.1.** *The DKSV Balanced-Separator instances have optimum value* $\Omega(\frac{\log\log n}{\log n})$.

(Although we haven't formally verified it, it's very likely that the optimum value of these instances is also $O(\frac{\log\log n}{\log n})$, at least for infinitely many $N$. The reason is that there is a $\sigma$-invariant function $f : \mathcal{F} \to \{-1,1\}$ of constant variance and total influence $\Omega(\log N)$; namely, $f(x,y) = 1$ if $x \in \{-1,1\}^N$ contains a "run" (with wraparound) of length $\lfloor \log_2 N - \log_2 \log\log N \rfloor$.)

On the other hand, the main result of [12] is the following:

**Theorem 6.2.** *The standard SDP relaxation with triangle inequalities for the DKSV Balanced-Separator instances has value* $O(\frac{1}{\log n})$.

We show here that this factor $\Theta(\log\log n)$ gap is eliminated when the degree-4 SOS relaxation is used.

**Theorem 6.3.** *The degree-4 SOS relaxation for the DKSV Balanced-Separator instances has value* $\Omega(\frac{\log\log n}{\log n})$.

*Proof.* We need to show

$$\{f(\mathcal{O})^2 = 1 : \forall \mathcal{O} \in V\} \cup \left\{ \left(\frac{1}{n}\sum_{\mathcal{O}\in V} f(\mathcal{O})\right)^2 \le \frac{1}{9},\ \frac{1}{|E|}\sum_{(\mathcal{O},\mathcal{O}')\in E} \left(\frac{f(\mathcal{O})-f(\mathcal{O}')}{2}\right)^2 \le c\frac{\log\log n}{\log n} \right\} \vdash_4 -1 \ge 0 \tag{12}$$

for some constant $c > 0$ (and $N$ sufficiently large).

Introduce indeterminates $g(x)$ for all $x \in \mathcal{F} = \{-1,1\}^N \times \{-1,1\}^N$. By Corollary 5.5 it is possible to write

$$-1 = u_0 + \sum_{x\in\mathcal{F}} v_x(g(x)^2-1) + \sum_{\substack{1\le i<j\le N\\ N+1\le i<j\le 2N}} w_{ij}(\mathbf{Inf}_i[g]-\mathbf{Inf}_j[g]) + u_1(\mathbf{Var}[g]-\tfrac{3}{4}) + u_2(\tfrac{1}{20}\ln(2N)-\mathbf{I}[g]), \tag{13}$$

where $u_0$, $u_1$, $u_2$ are SOS (in the variables $g(x)$) and all summands have degree at most 4. Now substitute into this identity $g(x) = f(\mathcal{O})$ for each $x \in \mathcal{O} \in V$, and also substitute $g(x) = 1$ for each $x \in \mathcal{F}$ which is not contained in any $\mathcal{O} \in V$. We now consider what happens to each term in (13).

First, we notice that the degree of each term cannot increase. The polynomial $u_0$ (now over indeterminates $f(\mathcal{O})$) remains SOS. The next term, $\sum_{x\in\mathcal{F}} v_x(g(x)^2-1)$, becomes of the form $\sum_{\mathcal{O}\in V} v'_{\mathcal{O}}(f(\mathcal{O})^2-1)$ for some polynomials $v'_{\mathcal{O}}$. We claim that each summand $w_{ij}(\mathbf{Inf}_i[g]-\mathbf{Inf}_j[g])$ in the next term drops out entirely. This is because when $g$ is viewed as mapping from $\mathcal{F}$ to the set of homogeneous degree-1 polynomials in the $f(\mathcal{O})$'s, it is invariant under the action of $\sigma$, by construction. From this it follows that $\mathbf{Inf}_i[g] = \mathbf{Inf}_j[g]$ formally as polynomials for all $1 \le i < j \le N$ and $N+1 \le i < j \le 2N$.

Next we come to the term $u_1(\mathbf{Var}[g] - \frac{3}{4})$. We have

$$\mathbf{Var}[g] - \tfrac{3}{4} = \mathop{\mathbf{E}}_{\boldsymbol{x}\in\mathcal{F}}[g(\boldsymbol{x})^2] - \tfrac{3}{4} - \mathop{\mathbf{E}}_{\boldsymbol{x}\in\mathcal{F}}[g(\boldsymbol{x})]^2.$$

Even after our substitution, $\mathbf{E}_{\boldsymbol{x}\in\mathcal{F}}[g(\boldsymbol{x})^2] - \frac{3}{4}$ will provably equal $\frac{1}{4}$ under the assumption $\{f(\mathcal{O})^2 = 1 : \forall \mathcal{O} \in V\}$, so it remains to focus on

$$\mathop{\mathbf{E}}_{\boldsymbol{x}\in\mathcal{F}}[g(\boldsymbol{x})]^2 = \left(\epsilon + (1-\epsilon)\tfrac{1}{n}\sum_{\mathcal{O}\in V} f(\mathcal{O})\right)^2.$$

Recalling that $\vdash_2 (Y+Z)^2 \le 2Y^2 + 2Z^2$, we deduce

$$\left(\tfrac{1}{n}\sum_{\mathcal{O}\in V} f(\mathcal{O})\right)^2 \le \tfrac{1}{9} \quad \vdash_2 \quad \left(\epsilon + (1-\epsilon)\tfrac{1}{n}\sum_{\mathcal{O}\in V} f(\mathcal{O})\right)^2 \le 2\epsilon^2 + 2(1-\epsilon)^2\cdot\tfrac{1}{9} \le \tfrac{1}{4}$$

(for $N$ sufficiently large, since $\epsilon = O(1/N^2)$), as needed.

Finally we come to the term $u_2(\frac{1}{20}\ln(2N) - \mathbf{I}[g])$. Let $\epsilon'$ denote the fraction of hypercube edges in $\mathcal{F}$ which are incident on $L$; note that $\epsilon' \le 2\epsilon = O(1/N^2)$. After our substitution, we have

$$\begin{aligned}\tfrac{1}{20}\ln(2N) - \mathbf{I}[g] &= \tfrac{1}{20}\ln(2N) - (2N)\mathop{\mathbf{E}}_{\substack{\text{edge }(\boldsymbol{x},\boldsymbol{y})\\ \text{in }\mathcal{F}}}\left[\left(\tfrac{g(\boldsymbol{x})-g(\boldsymbol{y})}{2}\right)^2\right] \\ &= \tfrac{1}{20}\ln(2N) - (2N)(1-\epsilon')\cdot\tfrac{1}{|E|}\sum_{(\mathcal{O},\mathcal{O}')\in E}\left(\tfrac{f(\mathcal{O})-f(\mathcal{O}')}{2}\right)^2 - \epsilon'\cdot(*), \qquad (14)\end{aligned}$$

where $(*)$ is the average of a number of terms, some of which are $(\frac{1-1}{2})^2 = 0$ and some of which are of the form

$$\left(\tfrac{f(\mathcal{O}_i)-1}{2}\right)^2 = 1 + \tfrac{1}{2}(f(\mathcal{O}_i)^2 - 1) - \tfrac{1}{4}(f(\mathcal{O}_i)+1)^2.$$

The above shows that $\{f(\mathcal{O})^2 = 1 : \forall\mathcal{O}\in V\} \vdash_2 \left(\frac{f(\mathcal{O}_i)-1}{2}\right)^2 \le 1$. Hence

$$\begin{aligned}&\{f(\mathcal{O})^2 = 1 : \forall\mathcal{O}\in V\} \cup \left\{\tfrac{1}{|E|}\sum_{(\mathcal{O},\mathcal{O}')\in E}\left(\tfrac{f(\mathcal{O})-f(\mathcal{O}')}{2}\right)^2 \le c\tfrac{\log\log n}{\log n}\right\} \\ &\qquad\qquad \vdash_2 \quad (14) \ge \tfrac{1}{20}\ln(2N) - (2N)(1-\epsilon')\cdot c\tfrac{\log\log n}{\log n} - \epsilon',\end{aligned}$$

which is nonnegative for $c$ sufficiently small, since $N = \Theta(\log n)$ and $\epsilon' = O(1/N^2)$. Thus we have verified (12). □

# 7 SOS proofs of the CLT and the $\frac{2}{\pi}$ Theorem

## 7.1 An invariance theorem for polynomials of linear forms

**Theorem 7.1.** *(SOS proof of an Invariance Theorem for polynomials of linear forms.)*

*Let $a_1, \dots, a_n$ be indeterminates. For any real vector $z = (z_1, \dots, z_n)$, let $\ell(z)$ denote the homogeneous linear polynomial $\ell(z) = a_1 z_1 + \cdots + a_n z_n$. Then for any even integer $k \ge 4$ we have*

$$a_1^2 + \cdots + a_n^2 \le 1 \quad \vdash_{2k} \quad \mathop{\mathbf{E}}_{\boldsymbol{G}}[\ell(\boldsymbol{G})^k] - k^{O(k)}\sum_{i=1}^n a_i^4 \le \mathop{\mathbf{E}}_{\boldsymbol{x}}[\ell(\boldsymbol{x})^k] \le \mathop{\mathbf{E}}_{\boldsymbol{G}}[\ell(\boldsymbol{G})^k],$$

*where $\boldsymbol{G} = (\boldsymbol{G}_1, \dots, \boldsymbol{G}_n) \sim \mathrm{N}(0,1)^n$ and $\boldsymbol{x} \sim \{-1,1\}^n$ is uniform.*

**Remark 7.2.** It is easy to see that $\mathbf{E}_{\boldsymbol{x}}[\ell(\boldsymbol{x})^k] = \mathbf{E}_{\boldsymbol{G}}[\ell(\boldsymbol{G})^k]$ formally as polynomials for $k = 0, 1, 2, 3$, and any odd integer $k > 3$.

*Proof.* For each integer $0 \le i \le n$, define the polynomial

$$P_i = \mathbf{E}[\ell(\boldsymbol{G}_1, \cdots, \boldsymbol{G}_i, \boldsymbol{x}_{i+1}, \cdots, \boldsymbol{x}_n)^k].$$

We will show for each $1 \le i \le n$ that

$$a_1^2 + \cdots + a_n^2 \le 1 \quad \vdash_{2k} \quad P_i - k^{O(k)} a_i^4 \le P_{i-1} \le P_i. \tag{15}$$

The desired result then follows by summing over $i$. So fix $1 \le i \le n$ and write $\ell(z) = \ell'(z') + a_i z_i$, where $z' = (z_1, \ldots, z_{i-1}, z_{i+1}, z_n)$ and

$$\ell'(z') = a_1 z_1 + \cdots + a_{i-1} z_{i-1} + a_{i+1} z_{i+1} + \cdots + a_n z_n$$

does not depend on the indeterminate $a_i$. Denoting $\boldsymbol{Z}' = (\boldsymbol{G}_1, \ldots, \boldsymbol{G}_{i-1}, \boldsymbol{x}_{i+1}, \ldots, \boldsymbol{x}_n)$ we have

$$\begin{aligned} P_i - P_{i-1} &= \mathop{\mathbf{E}}_{\boldsymbol{Z}'}\left[\mathop{\mathbf{E}}_{\boldsymbol{G}_i, \boldsymbol{x}_i}[(\ell'(\boldsymbol{Z}') + a_i \boldsymbol{G}_i)^k - (\ell'(\boldsymbol{Z}') + a_i \boldsymbol{x}_i)^k]\right] \\ &= \sum_{j=2}^{k/2} \binom{k}{2j} \left((2j-1)!! - 1\right) a_i^{2j} \mathop{\mathbf{E}}_{\boldsymbol{Z}'}[\ell'(\boldsymbol{Z}')^{k-2j}] \end{aligned} \tag{16}$$

where we used $\mathbf{E}[\boldsymbol{G}_i^r] = \mathbf{E}[\boldsymbol{x}_i^r] = 0$ for $r$ odd and $\mathbf{E}[\boldsymbol{G}_i^r] = (r-1)!!$, $\mathbf{E}[\boldsymbol{x}_i^r] = 1$, for $r$ even. The above polynomial is evidently SOS, justifying the second inequality in (15). As for the first inequality in (15), we have

$$a_1^2 + \cdots + a_n^2 \le 1 \quad \vdash_{2j} \quad a_i^{2j} \le a_i^4 \tag{17}$$

for each $i \in [n]$ and $2 \le j \le k/2$ because

$$\begin{aligned} a_i^4 - a_i^{2j} &= (1 - a_i^2)(a_i^4 + a_i^6 + a_i^8 + \cdots + a_i^{2j-2}) \\ &= \left((1 - \sum_{i'=1}^{n} a_{i'}^2) + \sum_{i' \ne i} a_{i'}^2\right)(a_i^4 + a_i^6 + a_i^8 + \cdots + a_i^{2j-2}); \end{aligned}$$

and, we have

$$\begin{aligned} a_1^2 + \cdots + a_n^2 \le 1 \quad \vdash_{2(k-2j)} \quad \mathop{\mathbf{E}}_{\boldsymbol{Z}'}[\ell'(\boldsymbol{Z}')^{k-2j}] &\le k^{O(k)} \mathop{\mathbf{E}}_{\boldsymbol{Z}'}[\ell'(\boldsymbol{Z}')^2]^{k/2-j} \\ &= k^{O(k)}(a_1^2 + \cdots + a_n^2)^{k/2-j} \le k^{O(k)} \end{aligned} \tag{18}$$

by Corollary 4.7, the second inequality's SOS proof being

$$1 - (\textstyle\sum a_{i'}^2)^{k/2-j} = 1 + (\sum a_{i'}^2) + (\sum a_{i'}^2)^2 + (\sum a_{i'}^2)^3 + \cdots + (\sum a_{i'}^2)^{k/2-j-1}.$$

Combining (17) and (18) via Fact 2.4

$$a_1^2 + \cdots + a_n^2 \le 1 \quad \vdash_{2k} \quad a_i^{2j} \mathop{\mathbf{E}}_{\boldsymbol{Z}'}[\ell'(\boldsymbol{Z}')^{k-2j}] \le k^{O(k)} a_i^4.$$

Using this in (16), along with $\binom{k}{2j}((2j-1)!! - 1) \le k^{O(k)}$ for each $j$, yields the first inequality in (15), completing the proof. ☐

## 7.2 An SOS proof of the $\frac{2}{\pi}$ Theorem

We require the below technical lemma giving a polynomial approximator to the absolute-value function. The proof uses some standard methods in approximation theory and is deferred to Appendix A.

**Lemma 7.3.** *For any sufficiently small parameter $\delta > 0$, there exists a univariate, real, even polynomial $P(t) = Q(t^2)$ of degree at most $\widetilde{O}(1/\delta^2)$ such that:*

1. *$P(t) \geq |t|$ for all $t \in \mathbb{R}$;*
2. *$\mathbf{E}[P(\sigma \boldsymbol{g})] \leq \sqrt{\frac{2}{\pi}} \cdot \sigma + \delta \leq \left(\frac{1}{2}\sigma^2 + \frac{1}{\pi}\right) + \delta$ for all $0 \leq \sigma \leq 1$, where $\boldsymbol{g} \sim \mathrm{N}(0,1)$;*
3. *Each coefficient of $P$ is at most $2^{O(d)}$ in absolute value.*

It is not hard to show that among degree-2 polynomials $P(t)$ with $P(t) \geq |t|$, the lowest possible value of $\mathbf{E}[P(\boldsymbol{g})]$ is 1, achieved by $P(t) = \frac{1}{2} + \frac{1}{2}t^2$. Interestingly, this is also the lowest possible value even when degree-4 is allowed:

**Theorem 7.4.** *Suppose $P(t)$ is a univariate real polynomial of degree at most 4 satisfying $P(t) \geq |t|$ for all real $t$. Then $\mathbf{E}_{\boldsymbol{g} \sim \mathrm{N}(0,1)}[P(\boldsymbol{g})] \geq 1$.*

*Proof.* Replacing $P(t)$ by $\frac{1}{2}(P(t) + P(-t))$ if necessary, we may assume $P(t)$ is even; i.e., $P(t) = a + bt^2 + ct^4$ for some real $a, b, c$. For any $M > 0$ we have

$$
\begin{array}{llllll}
 & \frac{1}{M^2} \times (P(0) \geq 0) & + & \frac{M^2-3}{M^2-1} \times (P(1) \geq 1) & + & \frac{2}{(M^2-1)M^2} \times (P(M) \geq M) \\
\Rightarrow & \frac{1}{M^2} \times (a \geq 0) & + & \frac{M^2-3}{M^2-1} \times (a + b + c \geq 1) & + & \frac{2}{(M^2-1)M^2} \times \left(a + M^2 b + M^4 c \geq M\right) \\
\Rightarrow & a + b + 3c \geq 1 - \frac{2}{M(M+1)}. & & & &
\end{array}
$$

This completes the proof because $\mathbf{E}[P(\boldsymbol{g})] = a + b + 3c$ and $M$ may be arbitrarily large. □

**Remark 7.5.** Once we allow degree 6 it is possible to obtain a bound strictly smaller than 1. For example, $P(t) = .333 + .815t^2 - .136t^4 + .01t^6 \geq |t|$ pointwise, and $\mathbf{E}[P(\boldsymbol{g})] = .89$.

The following "$\frac{2}{\pi}$ Theorem", due to [26], is essentially the special case of the Majority Is Stablest Theorem in which $\rho \to 0^+$. We reproduce the proof.

**Theorem 7.6.** *Let $f : \{-1,1\}^n \to [-1,1]$ and assume $|\widehat{f}(i)| \leq \epsilon$ for all $i \in [n]$. Then $\sum_{i=1}^n \widehat{f}(i)^2 \leq \frac{2}{\pi} + O(\epsilon)$.*

*Proof.* Let $\ell : \{-1,1\}^n \to \mathbb{R}$ be $\ell(x) = \sum_{i=1}^n \widehat{f}(i) x_i$ and let $\sigma = \sqrt{\sum_{i=1}^n \widehat{f}(i)^2}$. Then

$$
\sigma^2 = \mathop{\mathbf{E}}_{\boldsymbol{x} \sim \{-1,1\}^n}[f(\boldsymbol{x})\ell(\boldsymbol{x})] \leq \mathop{\mathbf{E}}_{\boldsymbol{x}}[|\ell(\boldsymbol{x})|] \leq \sigma \mathop{\mathbf{E}}_{\boldsymbol{g} \sim \mathrm{N}(0,1)}[|\boldsymbol{g}|] + O(\sigma\epsilon) = \sigma\left(\sqrt{\tfrac{2}{\pi}} + O(\epsilon)\right),
$$

the inequality being Berry–Esseen. The result follows after dividing by $\sigma$ and squaring. □

**Theorem 7.7.** *(SOS proof of the Berry–Esseen Theorem with $\ell_1$ functional.)*

*Let $a_1, \dots, a_n$ be indeterminates, and for each $x \in \{-1,1\}^n$, let $f(x)$ be an indeterminate. Let*

$$
A = \{f(x) \geq -1, f(x) \leq 1 : \forall x\} \cup \{a_1^2 + \cdots + a_n^2 \leq 1\}.
$$

*Then for any small real $\delta > 0$,*

$$A \quad \vdash_{\widetilde{O}(1/\delta^2)} \quad \mathop{\mathbf{E}}_{\boldsymbol{x}\sim\{-1,1\}^n}[f(\boldsymbol{x})(a_1\boldsymbol{x}_1 + \cdots + a_n\boldsymbol{x}_n)] \le b + \delta + 2^{\widetilde{O}(1/\delta^2)} \sum_{i=1}^n a_i^4,$$

*where we may choose either*

$$b = \sqrt{\tfrac{2}{\pi}} \qquad \text{or} \qquad b = \tfrac{1}{2}(a_1^2 + \cdots + a_n^2) + \tfrac{1}{\pi}.$$

*Proof.* For each $x \in \{-1,1\}^n$, let $\ell(x)$ denote $a_1x_1 + \cdots + a_nx_n$, a homogeneous linear polynomial in the indeterminates $a_i$. Let $P(t) = Q(t^2) = \sum_{k=0,2,4,\ldots,d} c_k t^k$ be the univariate real polynomial in Lemma 7.3, where $d = \deg(P) \le \widetilde{O}(1/\delta^2)$. Since $P(t) \ge \pm t$ for all real $t$, Fact 3.1 tells us that $\vdash_d P(t) \ge \pm t$ in $\mathbb{R}[t]$. Using Fact 3.6 and substituting $t = \ell(x)$ we deduce

$$\{f(x) \ge -1, f(x) \le 1\} \quad \vdash_{d+1} \quad f(x)\ell(x) \le P(\ell(x)).$$

Averaging over $x$ yields

$$\{f(x) \ge -1, f(x) \le 1 : \forall x\} \quad \vdash_{d+1} \quad \mathop{\mathbf{E}}_{\boldsymbol{x}}[f(\boldsymbol{x})(a_1\boldsymbol{x}_1 + \cdots + a_n\boldsymbol{x}_n)] \le \mathop{\mathbf{E}}_{\boldsymbol{x}}[P(\ell(\boldsymbol{x}))] = \sum_{k=0,2,4,\ldots,d} \mathop{\mathbf{E}}_{\boldsymbol{x}}[c_k\ell(\boldsymbol{x})^k]. \tag{19}$$

For each even $0 \le k \le d$, regardless of the sign of $c_k$, Theorem 7.1 implies that

$$a_1^2 + \cdots + a_n^2 \le 1 \quad \vdash_{2d} \quad \mathop{\mathbf{E}}_{\boldsymbol{x}}[c_k\ell(\boldsymbol{x})^k] \le \mathop{\mathbf{E}}_{\boldsymbol{G}}[c_k\ell(\boldsymbol{G})^k] + |c_k|k^{O(k)} \sum_{i=1}^n a_i^4.$$

Summing this over $k$, using $\sum_k |c_k|k^{O(k)} \le d^{O(d)}$, and combining with (19) yields

$$A \quad \vdash_{2d} \quad \mathop{\mathbf{E}}_{\boldsymbol{x}}[f(\boldsymbol{x})(a_1\boldsymbol{x}_1 + \cdots + a_n\boldsymbol{x}_n)] \le \mathop{\mathbf{E}}_{\boldsymbol{G}}[P(\ell(\boldsymbol{G}))] + d^{O(d)} \sum_{i=1}^n a_i^4. \tag{20}$$

Let $\sigma^2$ be shorthand for $\sum_{i=1}^n a_i^2$. Note that if we treat $a_1, \ldots, a_n$ as arbitrary *real numbers*, we have

$$\mathop{\mathbf{E}}_{\boldsymbol{G}}[P(\ell(\boldsymbol{G}))] = \mathop{\mathbf{E}}_{\boldsymbol{g}\sim\mathrm{N}(0,1)}[P(\sigma\boldsymbol{g})] = \mathop{\mathbf{E}}_{\boldsymbol{g}\sim\mathrm{N}(0,1)}[Q(\sigma^2\boldsymbol{g}^2)], \tag{21}$$

by the rotational symmetry of multivariate Gaussians. Since the left and right sides are polynomials in $a_1, \ldots, a_n$, it follows that (21) also holds as a formal polynomial identity over the indeterminates $a_1, \ldots, a_n$. Now temporarily view $\sigma^2$ as an indeterminate. From Lemma 7.3 we have that $\mathbf{E}_{\boldsymbol{g}\sim\mathrm{N}(0,1)}[Q(\sigma^2\boldsymbol{g}^2)]$ is upper-bounded by both $\sqrt{\frac{2}{\pi}} + \delta$ and $\frac{1}{2}\sigma^2 + \frac{1}{\pi} + \delta$ for all *real numbers* $0 \le \sigma^2 \le 1$. Thus from Fact 3.2 we have the following univariate SOS proof(s):

$$(1-\sigma^2)\sigma^2 \ge 0 \quad \vdash_{d/2} \quad \mathop{\mathbf{E}}_{\boldsymbol{g}\sim\mathrm{N}(0,1)}[Q(\sigma^2\boldsymbol{g}^2)] \le \sqrt{\tfrac{2}{\pi}} + \delta,\ \tfrac{1}{2}\sigma^2 + \frac{1}{\pi} + \delta$$

(note that $Q$ has even degree). Letting $\sigma^2 = \sum_{i=1}^n a_i^2$ again, we deduce that for either choice of $b$,

$$\left(1 - \sum_{i=1}^n a_i^2\right)\left(\sum_{i=1}^n a_i^2\right) \ge 0 \quad \vdash_d \quad \mathop{\mathbf{E}}_{\boldsymbol{g}\sim\mathrm{N}(0,1)}[Q(\sigma^2\boldsymbol{g}^2)] \le b + \delta$$
$$\Leftrightarrow \qquad a_1^2 + \cdots + a_n^2 \le 1 \quad \vdash_d \quad \mathop{\mathbf{E}}_{\boldsymbol{G}}[P(\ell(\boldsymbol{G}))] \le b + \delta,$$

using (21) and the fact that $\sum_{i=1}^n a_i^2$ is already SOS. Combining this with (20) yields

$$A \quad \vdash_{2d} \quad \mathop{\mathbf{E}}_{\boldsymbol{x}}[f(\boldsymbol{x})(a_1\boldsymbol{x}_1 + \cdots + a_n\boldsymbol{x}_n)] \leq b + \delta + d^{O(d)} \sum_{i=1}^n a_i^4,$$

as needed. ☐

**Corollary 7.8.** *(SOS proof of the $\frac{2}{\pi}$ Theorem.)*

*For each $x \in \{-1,1\}^n$, let $f(x)$ be an indeterminate. Define $\widehat{f}(S)$ as usual and write $\widehat{f}(i) = \widehat{f}(\{i\})$ for short. Let*

$$A = \{f(x) \geq -1, f(x) \leq 1 : \forall x\}.$$

*Then for each small real $\delta > 0$,*

$$A \quad \vdash_{\widetilde{O}(1/\delta^2)} \quad \sum_{i=1}^n \widehat{f}(i)^2 \leq \tfrac{2}{\pi} + \delta + 2^{\widetilde{O}(1/\delta^2)} \sum_{i=1}^n \widehat{f}(i)^4,$$

$$A \cup \{\widehat{f}(i)^2 \leq \tau : \forall i \in [n]\} \quad \vdash_{\widetilde{O}(1/\delta^2)} \quad \sum_{i=1}^n \widehat{f}(i)^2 \leq \tfrac{2}{\pi} + \delta + 2^{\widetilde{O}(1/\delta^2)} \cdot \tau.$$

*Proof.* We wish to apply Theorem 7.7 with $a_i = \widehat{f}(i)$ for each $i \in [n]$. A standard proof shows that

$$\sum_{S \subseteq [n]} \widehat{f}(S)^2 = \mathop{\mathbf{E}}_{\boldsymbol{x} \sim \{-1,1\}^n}[f(\boldsymbol{x})^2]$$

and hence, using Fact 3.4,

$$\vdash_3 \quad \sum_{i=1}^n \widehat{f}(i)^2 \leq 1. \tag{22}$$

We may therefore employ Theorem 7.7 (with $\delta/2$ instead of $\delta$) to obtain

$$A \quad \vdash_{\widetilde{O}(1/\delta^2)} \quad \mathop{\mathbf{E}}_{\boldsymbol{x} \sim \{-1,1\}^n}[f(\boldsymbol{x})(\widehat{f}(1)\boldsymbol{x}_1 + \cdots + \widehat{f}(n)\boldsymbol{x}_n)] \leq \tfrac{1}{2}\sum_{i=1}^n \widehat{f}(i)^2 + \tfrac{1}{\pi} + \tfrac{\delta}{2} + 2^{\widetilde{O}(1/\delta^2)} \sum_{i=1}^n \widehat{f}(i)^4.$$

But

$$\mathop{\mathbf{E}}_{\boldsymbol{x} \sim \{-1,1\}^n}[f(\boldsymbol{x})(\widehat{f}(1)\boldsymbol{x}_1 + \cdots + \widehat{f}(n)\boldsymbol{x}_n)] = \sum_{i=1}^n \widehat{f}(i)^2$$

is a polynomial identity so we deduce

$$\begin{aligned} A \quad \vdash_{\widetilde{O}(1/\delta^2)} \quad & \sum_{i=1}^n \widehat{f}(i)^2 \leq \tfrac{1}{2}\sum_{i=1}^n \widehat{f}(i)^2 + \tfrac{1}{\pi} + \tfrac{\delta}{2} + 2^{\widetilde{O}(1/\delta^2)} \sum_{i=1}^n \widehat{f}(i)^4 \\ \Leftrightarrow \quad & \sum_{i=1}^n \widehat{f}(i)^2 \leq \tfrac{2}{\pi} + \delta + 2^{\widetilde{O}(1/\delta^2)} \sum_{i=1}^n \widehat{f}(i)^4, \end{aligned}$$

completing the first part of the proof. Now adding the assumptions $\widehat{f}(i)^2 \leq \tau$ easily yields

$$A \cup \{\widehat{f}(i)^2 \leq \tau : \forall i \in [n]\} \quad \vdash_4 \quad \sum_{i=1}^n \widehat{f}(i)^4 \leq \tau \sum_{i=1}^n \widehat{f}(i)^2 \leq \tau$$

using (22) again. The proof is complete. ☐

## 8 Analysis of the KV Max-Cut instances

We recall the Max-Cut problem: Given is an undirected weighted graph $G$ on vertex set $V$ in which the nonnegative edge weights sum to 1. We write $(\boldsymbol{x},\boldsymbol{y}) \sim E$ to denote that $(\boldsymbol{x},\boldsymbol{y})$ is a random edge chosen with probability equal to the edge weight. It is required to find a cut $S \subseteq V$ so as to maximize $\mathbf{Pr}_{(\boldsymbol{x},\boldsymbol{y})\sim E}[\boldsymbol{x} \in S, \boldsymbol{y} \notin S \text{ or vice versa}]$. The natural polynomial optimization formulation has an indeterminate $f(x)$ for each vertex $x \in V$:

$$\begin{aligned} \max \quad & \mathop{\mathbf{E}}_{(\boldsymbol{x},\boldsymbol{y})\sim E}[\tfrac{1}{2} - \tfrac{1}{2} f(\boldsymbol{x})f(\boldsymbol{y})] \\ \text{s.t.} \quad & f(x)^2 = 1 \quad \forall x \in V. \end{aligned}$$

Thus as discussed in Section 2, the degree-$d$ SOS SDP hierarchy will use binary search to compute the smallest $\beta$ for which

$$\{f(x)^2 = 1 : \forall x \in V\} \cup \left\{ \mathop{\mathbf{E}}_{(\boldsymbol{x},\boldsymbol{y})\sim E}[\tfrac{1}{2} - \tfrac{1}{2} f(\boldsymbol{x})f(\boldsymbol{y})] \geq \beta \right\} \quad \vdash_d \quad -1 \geq 0.$$

**Unique-Games.** The Khot–Vishoi (KV) instances of Max-Cut [29] are given by composing the KKMO "noise stability" reduction from [26] with the KV integrality gap instances for Unique-Games (UG). Our SOS proof of the $\frac{2}{\pi}$ Theorem gives us a "black-box" analysis of the KKMO reduction which can essentially be "plugged in" to a sufficiently strong SOS analysis of UG instances. Let us now recall the Unique-Games problem with label-size $k \in \mathbb{N}^+$. Given is a regular weighted graph $\mathcal{G} = (\mathcal{V}, \mathcal{E})$ (self-loops allowed) with weights summing to 1. Also, given for each edge $(u, v)$ is a permutation $\pi_{uv} : [k] \to [k]$. We write $(\boldsymbol{u},\boldsymbol{v},\boldsymbol{\pi}) \sim \mathcal{E}$ to denote that edge $(\boldsymbol{u},\boldsymbol{v})$ with permutation $\boldsymbol{\pi} = \pi_{\boldsymbol{uv}}$ is chosen with probability equal to its edge weight. The goal is to give a labeling $F : \mathcal{V} \to [k]$ so as to maximize $\mathbf{Pr}_{(\boldsymbol{u},\boldsymbol{v},\boldsymbol{\pi})\sim\mathcal{E}}[\boldsymbol{\pi}(F(\boldsymbol{u})) = F(\boldsymbol{v})]$. The natural polynomial optimization formulation has an indeterminate $X_{u,i}$ for each $u \in \mathcal{V}, i \in [k]$:

$$\begin{aligned} \max \quad & \mathop{\mathbf{E}}_{(\boldsymbol{u},\boldsymbol{v},\boldsymbol{\pi})\sim\mathcal{E}}\left[\sum_{i=1}^k X_{\boldsymbol{u},i} X_{\boldsymbol{v},\boldsymbol{\pi}(i)}\right] = \mathop{\mathbf{E}}_{\boldsymbol{u}\in\mathcal{V}}\left[\sum_{i=1}^k X_{\boldsymbol{u},i} \cdot \mathop{\mathbf{E}}_{(\boldsymbol{v},\boldsymbol{\pi})\sim\boldsymbol{u}}[X_{\boldsymbol{v},\boldsymbol{\pi}(i)}]\right] \\ \text{s.t.} \quad & X_{u,i}^2 = X_{u,i} \quad \forall u \in \mathcal{V}, i \in [k] \\ & \sum_{i=1}^k X_{u,i} = 1 \quad \forall u \in \mathcal{V}, \end{aligned}$$

where we write $(\boldsymbol{v},\boldsymbol{\pi}) \sim u$ in place of $(u,\boldsymbol{v},\boldsymbol{\pi}) \sim \mathcal{E}|_{\boldsymbol{u}=u}$ for brevity. Thus the degree-$d$ SOS SDP hierarchy will use binary search to compute the smallest $\beta$ for which

$$\begin{aligned} &\{X_{u,i}^2 = X_{u,i} : \forall u \in \mathcal{V}, i \in [k]\} \cup \{\sum_{i=1}^k X_{u,i} = 1 : \forall u \in \mathcal{V}\} \\ &\qquad \cup \left\{ \mathop{\mathbf{E}}_{\boldsymbol{u}\in\mathcal{V}}\left[\sum_{i=1}^k X_{\boldsymbol{u},i} \cdot \mathop{\mathbf{E}}_{(\boldsymbol{v},\boldsymbol{\pi})\sim\boldsymbol{u}}[X_{\boldsymbol{v},\boldsymbol{\pi}(i)}]\right] \geq \beta \right\} \quad \vdash_d \quad -1 \geq 0. \end{aligned}$$

Barak et al. [3] have shown that the degree-4 moment SDP proves that the KV family of UG instances has a very low optimum value. In fact they show something stronger; one only needs the hypotheses $X_{u,i}^2 \leq X_{u,i}$ and $(\mathrm{avg}_{u,i} X_{u,i})^2 \leq 1/k^2$. Let us make a somewhat more general definition which applies to SOS-refutations of any UG instances:

**Definition 8.1.** Given a UG instance $\mathcal{G} = (\mathcal{V}, \mathcal{E})$ with label-size $k$, we say there is a *degree-$d$ SOS refutation that the* fractional assignment *optimum is at least $\beta$* if

$$\{X_{u,i} \geq 0 : \forall u \in \mathcal{V}, i \in [k]\} \cup \{\sum_{i=1}^{k} X_{u,i} \leq 1 : \forall u \in \mathcal{V}\}$$
$$\cup \left\{ \mathop{\mathbf{E}}_{\boldsymbol{u} \in \mathcal{V}} \Big[ \sum_{i=1}^{k} X_{\boldsymbol{u},i} \cdot \mathop{\mathbf{E}}_{(\boldsymbol{v},\boldsymbol{\pi}) \sim \boldsymbol{u}} [X_{\boldsymbol{v},\boldsymbol{\pi}(i)}] \Big] \geq \beta \right\} \quad \vdash_d \quad -1 \geq 0.$$

The above definition is slightly more demanding than the most natural one, in which the hypotheses $X_{u,i}^2 = X_{u,i}$ are granted. As mentioned, Barak et al. establish something noticeably stronger anyway: the following theorem is essentially proved in [3, Theorem 6.6]:[4]

**Theorem 8.2.** *Let $\mathcal{G} = \mathcal{G}(N, \eta) = (\mathcal{V}, \mathcal{E})$ be the Khot–Vishnoi instance of Unique-Games parameterized by $N$ (a power of 2) and $\eta \in (0, 1)$, which has $2^N/N$ vertices, label-size $N$, and optimum value at most $N^{-\eta}$. Then there is a degree-4 SOS refutation that its fractional assignment optimum is at least $N^{-\Omega(\eta)}$.*

We now recall the KKMO [26] reduction from UG to Max-Cut, which is parameterized by $\rho \in (-1, 0)$. Given a UG instance $\mathcal{G}$ with label-size $N$, the reduction creates a vertex set $V$ with a vertex $w_{u,x}$ for each $u \in \mathcal{V}$ and each $x \in \{-1, 1\}^N$. The probability distribution $E$ on edges for the Max-Cut instance is given as follows:

- draw $\boldsymbol{u} \sim \mathcal{V}$;
- independently draw $(\boldsymbol{u}, \boldsymbol{v}_1, \boldsymbol{\pi}_1)$ and $(\boldsymbol{u}, \boldsymbol{v}_2, \boldsymbol{\pi}_2)$ from the marginal of $\mathcal{E}$ which has first vertex $\boldsymbol{u}$;
- draw "$\rho$-correlated strings" $(\boldsymbol{x}, \boldsymbol{y})$ from $\{-1, 1\}^N$;
- output the edge $(w_{\boldsymbol{u}_1, \boldsymbol{x} \circ \boldsymbol{\pi}_1}, w_{\boldsymbol{u}_2, \boldsymbol{y} \circ \boldsymbol{\pi}_2})$.

KKMO make the following easy observation:

**Proposition 8.3.** *Consider any cut $V \to \{-1, 1\}$ in the above-described Max-Cut instance $(V, E)$; specifically, let us write it as a collection of functions $f_v : \{-1, 1\}^N \to \{-1, 1\}$, one for each $v \in \mathcal{V}$. Then the value of this cut is*

$$\tfrac{1}{2} - \tfrac{1}{2} \mathop{\mathbf{E}}_{\boldsymbol{u} \sim \mathcal{V}} [\mathbf{Stab}_\rho[g_{\boldsymbol{u}}]],$$

*where $g_u : \{-1, 1\}^N \to [-1, 1]$ is defined by $g_u(x) = \mathop{\mathbf{E}}_{(u, \boldsymbol{v}, \boldsymbol{\pi}) \sim \mathcal{E}|_{\boldsymbol{u}=u}} [f_{\boldsymbol{v}}(x \circ \boldsymbol{\pi})]$.*

As mentioned, the KV Max-Cut instances are formed by composing the KKMO reduction with the KV UG instances. Khot and Vishnoi show that for any fixed $\eta \in (0, 1)$, the optimum value of the resulting Max-Cut instance is at most $(\arccos \rho)/\pi + o_N(1)$. Further, using "Majority cuts" it's easy to show (using, e.g. [45, Theorem 3.4.2]) that the optimum values is at *least* $(\arccos \rho)/\pi - o_N(1)$.

The main result of this section is the following:

**Theorem 8.4.** *Fix any small $\epsilon, \delta > 0$. Let $\mathcal{G} = (\mathcal{V}, \mathcal{E})$ be a UG instance with label-size $N$ for which there is a degree-$d$ SOS proof ($d \geq 2$) that its fractional assignment optimum is at most $\epsilon$. Let $G = (V, E)$ be the Max-Cut instance resulting from applying the KKMO reduction with parameter $\rho \in (-1, 0)$ to $\mathcal{G}$. Then there is a degree $d + \tilde{O}(1/\delta^2)$ SOS refutation of the claim that the optimum value of $G$ is at least $\frac{1}{2} - \frac{1}{\pi}\rho - (\frac{1}{2} - \frac{1}{\pi})\rho^3 + \delta + \epsilon \cdot 2^{\tilde{O}(1/\delta^2)}$.*

[4] Take $k = 2$ therein, in which case Lemma 6.2 is obviated. This still only proves that optimum value of the degree-4 moment SDP is small. To get the fact that the optimum value is of the degree-4 SOS SDP is small (and hence that there is a refutation), one can either argue that there is no duality gap using ideas from the footnote in Section 2; or, one can use ideas from our proof of Theorem 5.1 to reprove their result from the SOS side.

Together with Theorem 8.2 this implies:

**Corollary 8.5.** *Fix any small $\delta > 0$. Let $\mathcal{G} = (\mathcal{V}, \mathcal{E})$ be a Khot-Vishnoi UG instance with label-size $N$ and noise parameter $\eta$. Let $G = (V, E)$ be the Max-Cut instance resulting from applying the KKMO reduction with parameter $\rho \in (-1, 0)$ to $\mathcal{G}$. Then there is a degree $\tilde{O}(1/\delta^2)$ SOS refutation of the claim that the optimum value of $G$ is at least $\frac{1}{2} - \frac{1}{\pi}\rho - (\frac{1}{2} - \frac{1}{\pi})\rho^3 + \delta + 2^{\tilde{O}(1/\delta^2)} \cdot N^{-\Omega(\eta)}$.*

**Corollary 8.6.** *Consider the KV Max-Cut instances with parameter $\rho_0 \approx -.689$. The degree-$O(1)$ SOS SDP certifies they have value at most $.779$, which is within a factor $.952$ of the optimum. For general $\rho$, the degree-$O(1)$ SOS SDP certifies a value for the KV Max-Cut instances which is within a factor $.931$ of the optimum, where*

$$.931 \approx \min_{\rho\in(-1,0)} \frac{(\arccos\rho)/\pi}{\frac{1}{2} - \frac{1}{\pi}\rho - (\frac{1}{2} - \frac{1}{\pi})\rho^3}.$$

Before proving Theorem 8.4 we prove a lemma which gives an alternative way to refute a UG instance having a good solution: roughly, for most vertices $v \in \mathcal{V}$, its neighbors cannot agree well on what $v$'s label should be.

**Lemma 8.7.** *Let $\mathcal{G} = (\mathcal{V}, \mathcal{E})$ be a UG instance with label-size $N$ and suppose there is a degree-$d$ SOS refutation that its fractional assignment optimum is at least $\epsilon$. Then also*

$$A \cup \left\{ \mathop{\mathbf{E}}_{\boldsymbol{u}\sim\mathcal{V}} \left[ \sum_{i=1}^{N} \Big( \mathop{\mathbf{E}}_{(\boldsymbol{v},\boldsymbol{\pi})\sim\boldsymbol{v}} [X_{\boldsymbol{v},\boldsymbol{\pi}(i)}] \Big)^2 \right] \geq 4\epsilon \right\} \quad \vdash_d \quad -1 \geq 0,$$

*where*

$$A = \{X_{u,i} \geq 0 : \forall x \in \mathcal{V}, i \in [N]\} \cup \{\sum_{i=1}^{N} X_{u,i} \leq 1 : \forall x \in \mathcal{V}\}.$$

*Proof.* Given the indeterminates $X_{u,i}$, define for each $u \in \mathcal{V}$ and $i \in [N]$ the homogeneous linear forms

$$Y_{u,i} = \tfrac{1}{2}X_{u,i} + \tfrac{1}{2} \mathop{\mathbf{E}}_{(\boldsymbol{v},\boldsymbol{\pi})\sim u} [X_{\boldsymbol{v},\boldsymbol{\pi}(i)}].$$

We will apply the assumption regarding the degree-$d$ SOS refutation for $\mathcal{G}$ to the $Y_{u,i}$'s. Certainly we have

$$A \quad \vdash_1 \quad Y_{u,i} \geq 0, \sum_{j=1}^{N} Y_{u,j} \leq 1$$

for every $u \in \mathcal{V}$, $i \in [N]$. Indeed, it's not hard to check that to complete the proof we need only verify

$$A \quad \vdash_2 \quad \mathop{\mathbf{E}}_{\boldsymbol{u}\sim\mathcal{V}} \left[ \sum_{i=1}^{N} Y_{\boldsymbol{u},i} \cdot \mathop{\mathbf{E}}_{(\boldsymbol{v},\boldsymbol{\pi})\sim\boldsymbol{u}} [Y_{\boldsymbol{v},\boldsymbol{\pi}(i)}] \right] \geq \tfrac{1}{4} \mathop{\mathbf{E}}_{\boldsymbol{u}\sim\mathcal{V}} \left[ \sum_{i=1}^{N} \Big( \mathop{\mathbf{E}}_{(\boldsymbol{v},\boldsymbol{\pi})\sim\boldsymbol{v}} [X_{\boldsymbol{v},\boldsymbol{\pi}(i)}] \Big)^2 \right].$$

But this follows from

$$\begin{aligned}
&\mathop{\mathbf{E}}_{\boldsymbol{u}} \left[ \sum_i Y_{\boldsymbol{u},i} \cdot \mathop{\mathbf{E}}_{(\boldsymbol{v},\boldsymbol{\pi})\sim\boldsymbol{u}} [Y_{\boldsymbol{v},\boldsymbol{\pi}(i)}] \right] \\
&= \tfrac{1}{2} \mathop{\mathbf{E}}_{\boldsymbol{u}} \left[ \sum_i X_{\boldsymbol{u},i} \cdot \mathop{\mathbf{E}}_{(\boldsymbol{v},\boldsymbol{\pi})\sim\boldsymbol{u}} [Y_{\boldsymbol{v},\boldsymbol{\pi}(i)}] \right] + \tfrac{1}{2} \mathop{\mathbf{E}}_{\boldsymbol{u}} \left[ \sum_i \mathop{\mathbf{E}}_{(\boldsymbol{v},\boldsymbol{\pi})\sim\boldsymbol{u}} [X_{\boldsymbol{v},\boldsymbol{\pi}(i)}] \cdot \mathop{\mathbf{E}}_{(\boldsymbol{v},\boldsymbol{\pi})\sim\boldsymbol{u}} [Y_{\boldsymbol{v},\boldsymbol{\pi}(i)}] \right] \\
&= \tfrac{1}{2} \mathop{\mathbf{E}}_{\boldsymbol{u}} \left[ \sum_i X_{\boldsymbol{u},i} \cdot \mathop{\mathbf{E}}_{(\boldsymbol{v},\boldsymbol{\pi})\sim\boldsymbol{u}} [Y_{\boldsymbol{v},\boldsymbol{\pi}(i)}] \right] \\
&\qquad + \tfrac{1}{4} \mathop{\mathbf{E}}_{\boldsymbol{u}} \left[ \sum_i \mathop{\mathbf{E}}_{(\boldsymbol{v},\boldsymbol{\pi})\sim\boldsymbol{u}} [X_{\boldsymbol{v},\boldsymbol{\pi}(i)}] \cdot \mathop{\mathbf{E}}_{(\boldsymbol{v}',\boldsymbol{\pi}')\sim\boldsymbol{v}} [X_{\boldsymbol{v}',\boldsymbol{\pi}'(i)}] \right] + \tfrac{1}{4} \mathop{\mathbf{E}}_{\boldsymbol{u}} \left[ \sum_i \Big( \mathop{\mathbf{E}}_{(\boldsymbol{v},\boldsymbol{\pi})\sim\boldsymbol{u}} [X_{\boldsymbol{v},\boldsymbol{\pi}(i)}]^2 \Big) \right]
\end{aligned}$$

(where we do not even need the assumptions $\sum_{i=1}^{N} X_{u,i} \leq 1$). ☐

We now give the proof of Theorem 8.4.

*Proof.* It is not hard to deduce the following result from Corollary 7.8:

**Corollary 8.8.** *In the setting of Corollary 7.8, for any* $\rho \in (-1, 0)$ *we have*

$$A \quad \vdash_{\widetilde{O}(1/\delta^2)} \quad \mathbf{Stab}_\rho[f] \geq \tfrac{2}{\pi} \cdot \rho + (1 - \tfrac{2}{\pi}) \cdot \rho^3 - \delta - 2^{\widetilde{O}(1/\delta^2)} \cdot \sum_{i=1}^{n} \widehat{f}(i)^4.$$

It is also easy to check using Fact 3.3 that

$$\{f_v(x)^2 = 1 : \forall v \in V, x \in \{-1,1\}^N\} \quad \vdash_4 \quad g_v(x) \geq -1, g_v(x) \leq 1$$

for all $v \in V$, $x \in \{-1,1\}^N$. Thus using the above corollary and Proposition 8.3 we obtain

$$\{f_v(x)^2 = 1 : \forall v \in V, x \in \{-1,1\}^N\}$$
$$\vdash_{\widetilde{O}(1/\delta^2)} \quad \tfrac{1}{2} - \tfrac{1}{2} \mathop{\mathbf{E}}_{\boldsymbol{u} \sim \mathcal{V}}[\mathbf{Stab}_\rho[g_{\boldsymbol{u}}]] \leq \tfrac{1}{2} - \tfrac{1}{\pi}\rho - (\tfrac{1}{2} - \tfrac{1}{\pi})\rho^3 + \delta + 2^{\widetilde{O}(1/\delta^2)} \cdot \mathop{\mathbf{E}}_{\boldsymbol{u} \sim \mathcal{V}}\Big[\sum_{i=1}^{N} \widehat{g_{\boldsymbol{u}}}(i)^4\Big]. \quad (23)$$

Now we bound the error term $2^{\widetilde{O}(1/\delta^2)} \cdot \mathbf{E}_{\boldsymbol{u} \sim \mathcal{V}}[\sum_{i=1}^N \widehat{g_{\boldsymbol{u}}}(i)^4]$ as follows. Using the polynomial identity $\widehat{g_u}(i) = \mathbf{E}_{(\boldsymbol{v},\boldsymbol{\pi}) \sim u}[\widehat{f_{\boldsymbol{v}}}(\boldsymbol{\pi}(i))]$ together with Fact 3.12 and Fact 3.13, we have

$$\vdash_4 \quad \mathop{\mathbf{E}}_{\boldsymbol{u} \sim \mathcal{V}}\Big[\sum_{i=1}^{N} \widehat{g_{\boldsymbol{v}}}(i)^4\Big] \leq \mathop{\mathbf{E}}_{\boldsymbol{u} \sim \mathcal{V}}\left[\sum_{i=1}^{N}\Big(\mathop{\mathbf{E}}_{(\boldsymbol{v},\boldsymbol{\pi}) \sim \boldsymbol{u}}[\widehat{f_{\boldsymbol{v}}}(\boldsymbol{\pi}(i))^2]\Big)^2\right]. \quad (24)$$

On the other hand, it is easy to check that for all $i \in [N]$ and $v \in \mathcal{V}$, we have

$$\{f_v(x)^2 = 1 : \forall x \in \{-1,1\}^N\} \quad \vdash_2 \quad \widehat{f_v}(i)^2 \geq 0, \ \sum_{i=1}^{N} \widehat{f_v}(i)^2 \leq 1. \quad (25)$$

Since there is a degree-$d$ refutation for $\mathcal{G}$ having a fractional assignment of value at least $\epsilon$, implementing Lemma 8.7 with $X_{v,i} = \widehat{f_v}(i)^2$, we have

$$\{\widehat{f_u}(i)^2 \geq 0 : \forall x \in \mathcal{V}, i \in [N]\} \cup \{\sum_{i=1}^{N} \widehat{f_u}(i)^2 \leq 1 : \forall x \in \mathcal{V}\}$$
$$\cup \left\{\mathop{\mathbf{E}}_{\boldsymbol{u} \sim \mathcal{V}}\left[\sum_{i=1}^{N}\Big(\mathop{\mathbf{E}}_{(\boldsymbol{v},\boldsymbol{\pi}) \sim \boldsymbol{u}} \widehat{f_{\boldsymbol{v}}}(\boldsymbol{\pi}(i))^2\Big)^2\right] \geq 4\epsilon\right\} \quad \vdash_{d+2} \quad -1 \geq 0. \quad (26)$$

By Fact 2.3, (25) and (26) give

$$\{f_v(x)^2 = 1 : \forall v \in \mathcal{V}, x \in \{-1,1\}^N\} \cup \left\{\mathop{\mathbf{E}}_{\boldsymbol{u} \sim \mathcal{V}}\left[\sum_{i=1}^{N}\Big(\mathop{\mathbf{E}}_{(\boldsymbol{v},\boldsymbol{\pi}) \sim \boldsymbol{u}} \widehat{f_{\boldsymbol{v}}}(\boldsymbol{\pi}(i))^2\Big)^2\right] \geq 4\epsilon\right\} \quad \vdash_{d+4} \quad -1 \geq 0. \quad (27)$$

Combining (27) and (24), we get

$$\{f_v(x)^2 = 1 : \forall v \in \mathcal{V}, x \in \{-1,1\}^N\} \cup \left\{\mathop{\mathbf{E}}_{\boldsymbol{u} \sim \mathcal{V}}\Big[\sum_{i=1}^{N} \widehat{g_{\boldsymbol{v}}}(i)^4\Big] \geq 4\epsilon\right\} \quad \vdash_{d+2} \quad -1 \geq 0. \quad (28)$$

Finally, combining (28) and (23), we get

$$\{f_v(x)^2 = 1 : \forall v \in \mathcal{V}, x \in \{-1,1\}^N\}$$
$$\cup \{\tfrac{1}{2} - \tfrac{1}{2} \mathop{\mathbf{E}}_{\boldsymbol{u} \sim \mathcal{V}}[\mathbf{Stab}_\rho[g_{\boldsymbol{u}}]] \geq \tfrac{1}{2} - \tfrac{1}{\pi}\rho - (\tfrac{1}{2} - \tfrac{1}{\pi})\rho^3 + \delta + 2^{\widetilde{O}(1/\delta^2)}\} \quad \vdash_{d+\tilde{O}(1/\delta^2)} \quad -1 \geq 0. \quad \square$$

**Acknowledgments**

We thank the (other) authors of [3] for their insights, especially Boaz Barak, Jon Kelner, and David Steurer. We thank Paul Beame, Monique Laurent, Toni Pitassi, Ali Kemal Sinop, David Witmer, and John Wright for helpful discussions.

# References

[1] Emil Artin. Über die Zerlegung definiter Funktionen in Quadrate. *Abhandlungen aus dem Mathematischen Seminar der Universität Hamburg*, 5(1):100–115, 1927. 1.1

[2] Albert Atserias and Víctor Dalmau. A combinatorial characterization of resolution width. *Journal of Computer and System Sciences*, 74(3):323–334, 2008. 1

[3] Boaz Barak, Fernando Brandão, Aram Harrow, Jonathan Kelner, David Steurer, and Yuan Zhou. Hypercontractivity, sum-of-squares proofs, and their applications. In *Proceedings of the 44th Annual ACM Symposium on Theory of Computing*, pages 307–326, 2012. 1, 1.1, 1.2, 1.2, 1.2, 4, 4.3, 4, 5.1, 8, 8, 8

[4] Boaz Barak, Parikshit Gopalan, Johan Håstad, Raghu Meka, Prasad Raghavendra, and David Steurer. Making the long code shorter, with applications to the Unique Games Conjecture. In *Proceedings of the 53rd Annual IEEE Symposium on Foundations of Computer Science*, 2012. 1

[5] Paul Beame, Russell Impagliazzo, Jan Krajíček, Toniann Pitassi, and Pavel Pudlák. Lower bounds on Hilbert's Nullstellensatz and propositional proofs. *Proceedings of the London Mathematical Society*, 3(1):1–26, 1996. 1.1

[6] Jacek Bochnak, Michel Coste, and Marie-Françoise Roy. *Real algebraic geometry*. Springer, 1998. 1.1

[7] Aline Bonami. Étude des coefficients Fourier des fonctions de $L^p(G)$. *Annales de l'Institute Fourier*, 20(2):335–402, 1970. 1.2, 4

[8] Christer Borell. Geometric bounds on the Ornstein-Uhlenbeck velocity process. *Probability Theory and Related Fields*, 70(1):1–13, 1985. 1.2

[9] Sam Buss, Dima Grigoriev, Russell Impagliazzo, and Toniann Pitassi. Linear gaps between degrees for the polynomial calculus modulo distinct primes. In *Proceedings of the 31st Annual ACM Symposium on Theory of Computing*, pages 547–556, 1999. 1.1

[10] Kevin Cheung. Computation of the Lasserre ranks of some polytopes. *Mathematics of Operations Research*, 32(1):88–94, 2007. 1.1

[11] Matthew Clegg, Jeffrey Edmonds, and Russell Impagliazzo. Using the Groebner basis algorithm to find proofs of unsatisfiability. In *Proceedings of the 28th Annual ACM Symposium on Theory of Computing*, pages 174–183, 1996. 1

[12] Nikhil Devanur, Subhash Khot, Rishi Saket, and Nisheeth Vishnoi. Integrality gaps for Sparsest Cut and Minimum Linear Arrangement problems. In *Proceedings of the 38th Annual ACM Symposium on Theory of Computing*, pages 537–546, 2006. 1, 1.2, 6, 6

[13] Ilias Diakonikolas, Parikshit Gopalan, Ragesh Jaiswal, Rocco Servedio, and Emanuele Viola. Bounded independence fools halfspaces. *SIAM Journal on Computing*, 39(8):3441–3462, 2010. A, A

[14] Michel Goemans and David Williamson. Improved approximation algorithms for maximum cut and satisfiability problems using semidefinite programming. *Journal of the ACM*, 42:1115–1145, 1995. 1.2

[15] Dima Grigoriev. Linear lower bound on degrees of Positivstellensatz calculus proofs for the parity. Technical Report IHES/M/99/68, Insitut des Hautes Études Scientifiques, 1999. 1.1

[16] Dima Grigoriev. Complexity of Positivstellensatz proofs for the knapsack. *Computational Complexity*, 10(2):139–154, 2001. 1.1

[17] Dima Grigoriev. Linear lower bound on degrees of Positivstellensatz calculus proofs for the parity. *Theoretical Computer Science*, 259(1-2):613–622, 2001. 1.1

[18] Dima Grigoriev, Edward Hirsch, and Dmitrii Pasechnik. Complexity of semialgebraic proofs. *Moscow Mathematical Journal*, 2(4):647–679, 2002. 1.1

[19] Dima Grigoriev and Nicolai Vorobjov. Complexity of Null- and Positivstellensatz proofs. *Annals of Pure and Applied Logic*, 113(1):153–160, 2001. 1, 1.1

[20] Venkatesan Guruswami, Ali Sinop, and Yuan Zhou. Constant factor Lasserre integrality gaps for graph partitioning problems. Technical Report 1202.6071, arXiv, 2012. 1.1

[21] David Hilbert. Über die Darstellung definiter Formen als Summe von Formenquadraten. *Mathematische Annalen*, 32(3):342–350, 1888. 1.1

[22] David Hilbert. Mathematical problems. *Bulletin of the American Mathematical Society*, 8(10):437–479, 1902. 1.1

[23] Jeff Kahn, Gil Kalai, and Nathan Linial. The influence of variables on Boolean functions. In *Proceedings of the 29th Annual IEEE Symposium on Foundations of Computer Science*, pages 68–80, 1988. 1.2, 5.1, 5.2

[24] Anna Karlin, Claire Mathieu, and C. Thach Nguyen. Integrality gaps of linear and semi-definite programming relaxations for Knapsack. In *Proceedings of the 15th Annual Conference on Integer Programming and Combinatorial Optimization*, pages 301–314, 2011. 1.1

[25] Subhash Khot, Guy Kindler, Elchanan Mossel, and Ryan O'Donnell. Optimal inapproximability results for MAX-CUT and other 2-variable CSPs? In *Proceedings of the 45th Annual IEEE Symposium on Foundations of Computer Science*, pages 146–154, 2004. 1.2

[26] Subhash Khot, Guy Kindler, Elchanan Mossel, and Ryan O'Donnell. Optimal inapproximability results for Max-Cut and other 2-variable CSPs? *SIAM Journal on Computing*, 37(1):319–357, 2007. 1.2, 7.2, 8, 8

[27] Subhash Khot, Preyas Popat, and Rish Saket. Approximate Lasserre integrality gap for Unique Games. In *Proceedings of the 13th Annual International Workshop on Approximation Algorithms for Combinatorial Optimization Problems*, pages 298–311, 2010. 1

[28] Subhash Khot and Rishi Saket. SDP integrality gaps with local $\ell_1$-embeddability. In *Proceedings of the 50th Annual IEEE Symposium on Foundations of Computer Science*, pages 565–574, 2009. 1.2

[29] Subhash Khot and Nisheeth Vishnoi. The Unique Games Conjecture, integrality gap for cut problems and embeddability of negative type metrics into $\ell_1$. In *Proceedings of the 46th Annual IEEE Symposium on Foundations of Computer Science*, pages 53–62, 2005. 1, 1.2, 1.2, 8

[30] Guy Kindler and Ryan O'Donnell. Gaussian noise sensitivity and Fourier tails. In *Proceedings of the 26th Annual IEEE Conference on Computational Complexity*, 2012. 1.2

[31] Robert Krauthgamer and Yuval Rabani. Improved lower bounds for embeddings into $L_1$. *SIAM Journal on Computing*, 38(6):2487–2498, 2009. 1.2

[32] Jean-Louis Krivine. Anneaux préordonnés. *Journal d'Analyse Mathématique*, 12(1):307–326, 1964. 1, 1.1

[33] Jean Lasserre. Optimisation globale et théorie des moments. *Comptes Rendus de l'Académie des Sciences*, 331(11):929–934, 2000. 1, 1.1

[34] Jean Lasserre. Global optimization with polynomials and the problem of moments. *SIAM Journal on Optimization*, 11(3):796–817, 2001. 1, 1.1, 1.1

[35] Monique Laurent. Lower bound for the number of iterations in semidefinite relaxations for the cut polytope. *Mathematics of Operations Research*, 28(4):871–883, 2003. 1.1

[36] Monique Laurent. Semidefinite relaxations for max-cut. In Martin Grötschel, editor, *The Sharpest Cut*, chapter 16, pages 257–290. Society for Industrial and Applied Mathematics and the Mathematical Programming Society, 2004. 1.1

[37] Monique Laurent. Semidefinite representations for finite varieties. *Mathematical Programming*, 109(1):1–26, 2007. 1.1

[38] Monique Laurent. Sums of squares, moment matrices and optimization over polynomials. *Emerging Applications of Algebraic Geometry*, 149:157–270, 2009. 1.1, 3

[39] Henri Lombardi, Nikolai Mnev, and Marie-Fran coise Roy. The Positivstellensatz and small deduction rules for systems of inequalities. *Mathematische Nachrichten*, 181(1):245–259, 1996. 1

[40] Vladimir Andreevich Markov. On functions of least deviation from zero in a given interval. 1892. A

[41] Murray Marshall. *Positive polynomials and sums of squares*. American Mathematical Society, 2008. 2, 3

[42] Elchanan Mossel, Ryan O'Donnell, and Krzysztof Oleszkiewicz. Noise stability of functions with low influences: invariance and optimality. *Annals of Mathematics*, 171(1), 2010. 1.2, 4, 4

[43] Fedor Nazarov. http://mathoverflow.net/questions/97769/approximation-theory-reference-for-a-bounded-polynomial-having-bounded-coefficien, 2012. A

[44] Yurii Nesterov. Global quadratic optimization via conic relaxation. In *Handbook of Semidefinite Programming*, pages 363–384. Kluwer Academic Publishers, 2000. 1

[45] Ryan O'Donnell. *Computational applications of noise sensitivity*. PhD thesis, Massachusetts Institute of Technology, 2003. 8

[46] Ryan O'Donnell. Some topics in analysis of Boolean functions. In *Proceedings of the 40th Annual ACM Symposium on Theory of Computing*, pages 569–578, 2008. 4

[47] Pablo Parrilo. *Structured Semidefinite Programs and Semialgebraic Geometry Methods in Robustness and Optimization*. PhD thesis, California Institute of Technology, 2000. 1, 1.1

[48] Toniann Pitassi. Propositional proof complexity and unsolvability of polynomial equations. In *Proceedings of International Congress of Mathematicians Meeting*, pages 451–460, 1998. 1.1

[49] Mihai Putinar. Positive polynomials on compact semi-algebraic sets. *Indiana University Mathematics Journal*, 42(3):969–984, 1993. 1, 1.1

[50] Prasad Raghavendra and David Steurer. Integrality gaps for strong SDP relaxations of Unique Games. In *Proceedings of the 50th Annual IEEE Symposium on Foundations of Computer Science*, pages 575–585, 2009. 1, 1.2

[51] Konrad Schmüdgen. The K-moment problem for compact semi-algebraic sets. *Mathematische Annalen*, 289(1):203–206, 1991. 1.1

[52] Alexander Sherstov. Making polynomials robust to noise. In *Proceedings of the 44th Annual ACM Symposium on Theory of Computing*, pages 747–758, 2012. A

[53] Naum Shor. An approach to obtaining global extremums in polynomial mathematical programming problems. *Cybernetics*, 23(5):695–700, 1987. 1.1

[54] Naum Shor. Class of global minimum bounds of polynomial functions. *Cybernetics*, 23(6):731–734, 1987. 1, 1.1

[55] Gilbert Stengle. A Nullstellensatz and a Positivstellensatz in semialgebraic geometry. *Mathematische Annalen*, 207(2):87–97, 1973. 1, 1.1

[56] Madhur Tulsiani. CSP gaps and reductions in the Lasserre hierarchy. In *Proceedings of the 41st Annual ACM Symposium on Theory of Computing*, pages 303–312, 2009. 1.1

[57] Thorsten Wörmann. *Strikt positive Polynome in der semialgebraischen Geometrie*. PhD thesis, TU Dortmund University, 1998. 1.1

# A An approximator for the absolute-value function

Here we restate and prove Lemma 7.3. A key tool will be the polynomial approximator to the $\mathrm{sgn}$ function constructed in [13].

**Lemma 7.3.** *For any sufficiently small parameter $\delta > 0$, there exists a univariate, real, even polynomial $P(t) = Q(t^2)$ of degree at most $\widetilde{O}(1/\delta^2)$ such that:*

1. $P(t) \geq |t|$ *for all* $t \in \mathbb{R}$*;*

2. $\mathbf{E}[P(\sigma \boldsymbol{g})] \leq \sqrt{\frac{2}{\pi}} \cdot \sigma + \delta \leq \left(\frac{1}{2}\sigma^2 + \frac{1}{\pi}\right) + \delta$ *for all* $0 \leq \sigma \leq 1$*, where* $\boldsymbol{g} \sim \mathrm{N}(0,1)$*;*

3. *Each coefficient of* $P$ *is at most* $2^{O(\deg(P))}$ *in absolute value.*

*Proof.* We will use the following result from [13, Theorem 3.10]:

**Theorem A.1.** *For every* $0 < \epsilon < .1$ *there is an odd integer* $d = d(\epsilon) = \Theta(\log^2(1/\epsilon)/\epsilon)$ *and a univariate polynomial* $p(t)$ *of degree* $d$ *satisfying:*

- $p(t) \in [\mathrm{sgn}(t) - \epsilon, \mathrm{sgn}(t) + \epsilon]$ *for all* $|t| \in [\epsilon, 1]$*;*
- $p(t) \in [-1 - \epsilon, 1 + \epsilon]$ *for all* $|t| \le \epsilon$*;*
- $p(t)$ *is monotonically increasing on the intervals* $(-\infty, -1]$ *and* $[1, +\infty)$.

We can assume without loss of generality that $p(t)$ is odd since the odd part of $p(t)$ (i.e. $(p(t) - p(-t))/2$) also satisfies the properties in Theorem A.1.

Given $p(t)$ as in Theorem A.1, define

$$p_0(t) = (1 + 2\epsilon)p(t/M), \quad \text{where } M = \frac{c \log^c(1/\epsilon)}{\sqrt{\epsilon}}$$

and $c > 1$ is a universal constant to be chosen later. The polynomial $p_0(t)$ has the following properties:

- $p_0(t) \in [1, 1 + 4\epsilon]$ when $t \in [M\epsilon, M]$, $p_0(t) \in [-(1 + 4\epsilon), -1]$ when $t \in [-M, -M\epsilon]$;
- $p_0(t) \in [-(1 + 4\epsilon), 1 + 4\epsilon]$ for all $|t| \le M\epsilon$;
- $p_0(t) \ge 1$ when $t \ge M$, $p_0(t) \le 1$ when $t \le -M$.

Finally, define

$$P(t) = \int_0^t p_0(x)dx + 2M\epsilon.$$

an even polynomial of degree $d + 1$. We will show that the following hold assuming $c$ is taken sufficiently large and then $\epsilon$ is sufficiently small:

(*a*) $P(t) \ge |t|$ for all $t \in \mathbb{R}$;

(*b*) $\mathbf{E}[P(\sigma \boldsymbol{g})] \le \sqrt{\frac{2}{\pi}} \cdot \sigma + O(M\epsilon)$ for all $0 \le \sigma \le 1$;

(*c*) Each coefficient of $P$ is at most $2^{O(d)}$ in absolute value.

The proof is then completed by taking $\epsilon = \delta^2/\mathrm{polylog}(1/\delta)$.

Properties (*a*) follows easily from the definition of $P(t)$. It also follows easily from the definition that $|P(t)| \le 1 + O(M\epsilon) \le 2$ for all $|t| \le 1$. It is a standard fact in approximation theory (see, e.g., [52, 43]) that if $P$ is a degree $d + 1$ polynomial satisfying $|P(t)| \le b$ for all $|t| \le 1$ then each coefficient of $P(t)$ is at most, say, $b(4e)^{d+1} = 2^{O(d)}$ in magnitude. This verifies (*c*). It remains to establish property (*b*). For this we have

$$\mathbf{E}[P(\sigma \boldsymbol{g})] = \mathbf{E}[P(\sigma \boldsymbol{g}) \cdot \mathbf{1}\{|\sigma \boldsymbol{g}| \le M\}] + \mathbf{E}[P(\sigma \boldsymbol{g}) \cdot \mathbf{1}\{|\sigma \boldsymbol{g}| > M\}] \tag{29}$$

Regarding the first term in (29) we use that for $|t| \le M$ we have $|p_0(t)| \le 1 + 4\epsilon$ and hence

$$P(t) \le (1 + 4\epsilon)|t| + 2M\epsilon = |t| + O(M\epsilon) \quad \forall |t| \le M. \tag{30}$$

Thus

$$\mathbf{E}[P(\sigma \boldsymbol{g}) \cdot \mathbf{1}\{|\sigma \boldsymbol{g}| \le M\}] \le \mathbf{E}[|\sigma \boldsymbol{g}| \cdot \mathbf{1}\{|\sigma \boldsymbol{g}| \le M\}] + O(M\epsilon) \le \mathbf{E}[|\sigma \boldsymbol{g}|] + O(M\epsilon) = \sqrt{\tfrac{2}{\pi}} \cdot \sigma + O(M\epsilon).$$

To complete the verification of (*b*) it therefore suffices to bound the second term in (29) by $O(M\epsilon)$. In fact we will show

$$\mathbf{E}[P(\sigma \boldsymbol{g}) \cdot \mathbf{1}\{|\sigma \boldsymbol{g}| > M\}] \ll M\epsilon. \tag{31}$$

Using evenness of $P$ and the fact that it is evidently increasing on $[M, \infty)$ we have

$$\mathbf{E}[P(\sigma \boldsymbol{g}) \cdot \mathbf{1}\{|\sigma \boldsymbol{g}| > M\}] = 2\,\mathbf{E}[P(\sigma \boldsymbol{g}) \cdot \mathbf{1}\{\sigma \boldsymbol{g} > M\}] \leq 2\,\mathbf{E}[P(\boldsymbol{g}) \cdot \mathbf{1}\{\boldsymbol{g} > M\}]. \tag{32}$$

We upper-bound $P$'s value on large inputs using a well-known fact from approximation theory (and a corollary of the theorem in §33 of [40]):

**Fact A.2.** *Let $q(t)$ be a polynomial of degree at most $k$ satisfying $|q(t)| \leq b$ for all $|t| \leq 1$. Then $|q(t)| \leq b\,|3t|^k$ for all $|t| \geq 1$.*

Applying this fact to $p(t)$ we obtain $p(t) \leq (1+\epsilon)(3t)^d$ for all $t \geq 1$, whence $p_0(t) \leq 2(3t/M)^d$ for all $t \geq M$, whence

$$P(t) = \int_0^t p_0(x)dx + 2M\epsilon \leq O(M) + t \cdot 2(3t/M)^d \leq O(1) \cdot (3t/M)^{d+1}$$

for all $t \geq M$ (we also used $M = o(2^d)$). Thus

$$\begin{aligned}(32) \leq O(\tfrac{3}{M})^{d+1} \cdot \mathbf{E}[\boldsymbol{g}^{d+1} \cdot \mathbf{1}\{\boldsymbol{g} > M\}] &\leq O(\tfrac{3}{M})^{d+1} \cdot O(dM)^{d+1} \exp(-M^2/2) \\ &\leq 2^{\text{polylog}(1/\epsilon)/\epsilon} \exp(-M^2/2) = 2^{\text{polylog}(1/\epsilon)/\epsilon} \exp(-c^2 \log^{2c}(1/\epsilon)/2\epsilon) \ll M\epsilon\end{aligned}$$

if we choose $c$ to be a large enough universal constant. This completes the justification of (31) and the overall proof. □